\documentclass[11pt,a4paper]{article}
\pdfoutput=1
\usepackage[english]{babel}
\usepackage[T1]{fontenc}
\usepackage[latin1]{inputenc}
\usepackage{amsfonts,amsbsy,bm,euscript,mathrsfs}
\usepackage{amssymb,stmaryrd,faktor,slashed,arydshln}
\usepackage[x11names]{xcolor}
\usepackage[tbtags]{amsmath}
\usepackage[linktocpage,bookmarks=true,colorlinks=true,linkcolor=black,citecolor=black,urlcolor=black,bookmarksnumbered]{hyperref}
\usepackage{cancel,url,verbatim,todonotes,breqn}
\usepackage[yyyymmdd,hhmmss]{datetime}

\usepackage[nosort]{cite}

\numberwithin{equation}{section}

\allowdisplaybreaks[3]

\makeatletter
\renewcommand\section{\@startsection {section}{1}{\z@}
{-3.5ex \@plus -1ex \@minus -.2ex}
{2.3ex \@plus.2ex}
{\normalfont\Large\bfseries}}
\renewcommand\subsection{\@startsection{subsection}{2}{\z@}
{-3.25ex\@plus -1ex \@minus -.2ex}
{1.5ex \@plus.2ex}
{\normalfont\large\bfseries}}
\def\maketag@@@#1{\hbox{\m@th\normalfont\normalsize#1}}
\makeatother

\RequirePackage{verbatim}

\makeatletter
\newwrite\bibinl@out

\makeatother

\expandafter\def\expandafter\bfseries\expandafter{\bfseries\ifmmode\else\boldmath\fi}
\expandafter\def\expandafter\mdseries\expandafter{\mdseries\ifmmode\else\unboldmath\fi}
\expandafter\def\expandafter\normalfont\expandafter{\normalfont\ifmmode\else\unboldmath\fi}

\newcommand \arxivlink [1]{\href{http://arxiv.org/abs/#1}{\tt arXiv:#1}}

\newcommand \Ad {\operatorname{Ad}}
\newcommand \Tr {\operatorname{Tr}}
\newcommand \diag {\operatorname{diag}}

\def\S{{\cal S}}
\def\C{{\cal C}}
\def\cF{{\cal F}}
\def\ww{\wedge}

\begin{document}

\begin{titlepage}
\thispagestyle{empty}
\begin{center}

\hfill

\vspace{1.5truecm}

{\LARGE \bf Marginal and non-commutative deformations \\ via non-abelian T-duality}

\vspace{1.5truecm}

{Ben Hoare$^{a}$ and Daniel C. Thompson$^{b}$}

\vspace{1.0truecm}

{\em $^{a}$ Institut f\"ur Theoretische Physik, ETH Z\"urich,\\ Wolfgang-Pauli-Strasse 27, 8093 Z\"urich, Switzerland.}

\vspace{0.25truecm}

{\em $^{b}$ Theoretische Natuurkunde, Vrije Universiteit Brussel \& The International Solvay Institutes, \\ Pleinlaan 2, B-1050 Brussels, Belgium.}

\vspace{0.5truecm}

{\em E-mail: \ } {\tt \href{mailto:bhoare@ethz.ch}{bhoare@ethz.ch}, \href{mailto:Daniel.Thompson@vub.ac.be}{Daniel.Thompson@vub.ac.be}}

\vspace{1.5truecm}

\end{center}

\begin{abstract}
In this short article we develop recent proposals to relate Yang-Baxter sigma-models and non-abelian T-duality. We demonstrate explicitly that the holographic space-times associated to both (multi-parameter)-$\beta$-deformations and non-commutative deformations of ${\cal N}=4$ super Yang-Mills gauge theory including the RR fluxes can be obtained via the machinery of non-abelian T-duality in Type II supergravity.
\end{abstract}

\vfill

\setcounter{footnote}{0}
\end{titlepage}

\section{Introduction}\label{sec:intro}

There is a rich interplay between the three ideas of T-duality, integrability and holography. Perhaps the most well studied example of this is the use of the TsT transformation to ascertain the gravitational dual space-times to certain marginal deformations of ${\cal N}=4$ super Yang-Mills gauge theory \cite{Lunin:2005jy}. Whilst this employs familiar T-dualities of $U(1)$ isometries in space-time, T-duality can be extended to both non-abelian isometry groups and to fermionic directions in superspace. Such generalised T-dualities also have applications to holography. Fermionic T-duality \cite{Berkovits:2008ic,Beisert:2008iq} was critical in understanding the scattering amplitude/Wilson loop duality at strong coupling. T-duality of non-abelian isometries has been employed as a solution generating technique in Type II supergravity \cite{Sfetsos:2010uq}, relating for instance $AdS_5\times S^5$ to (a limit\footnote{A more precise field theoretic explanation of what this limit means has been proposed in \cite{Lozano:2016kum}.} 
of) the space-times corresponding to ${\cal N}=2$ non-Lagrangian gauge theories. Developing the recent results of \cite{Hoare:2016wsk,Borsato:2016pas} this note will investigate further the role generalised notions of T-duality can play in holography.

A new perspective on deformations of the $AdS_5 \times S^5$ superstring has come from the study of Yang-Baxter deformations of string $\sigma$-models \cite{Klimcik:2002zj,Klimcik:2008eq,Klimcik:2014bta,Delduc:2013fga,Delduc:2013qra}. These are integrable algebraic constructions which deform the target space of the $\sigma$-model through the specification of an antisymmetric $r$-matrix solving the (modified) classical Yang-Baxter equation ((m)cYBE).

If the $r$-matrix solves the mcYBE then, applied to the supercoset formulation of strings in $AdS_5\times S^5$ \cite{Metsaev:1998it,Berkovits:1999zq}, these give rise to $\eta$-deformed space-times which are conjectured to encode a quantum group $q$-deformation of ${\cal N}=4$ super Yang-Mills with a deformation parameter $q \in \mathbb{R}$ \cite{Delduc:2014kha,Arutyunov:2013ega,Arutyunov:2015qva}. However the $\eta$-deformed worldsheet theory appears to be only globally scale invariant \cite{Hoare:2015gda,Hoare:2015wia}, the target space-time does not solve exactly the Type II supergravity equations \cite{Arutyunov:2015qva} but rather a generalisation thereof \cite{Arutyunov:2015mqj}. Classically $\eta$-deformations are related via a generalised Poisson-Lie T-duality \cite{Vicedo:2015pna,Hoare:2015gda,Sfetsos:2015nya,Klimcik:2015gba,Klimcik:2016rov,Delduc:2016ihq} to a class of integrable deformation of (gauged) WZW models known as $\lambda$-deformations \cite{Sfetsos:2013wia,Hollowood:2014rla,Hollowood:2014qma}, which do however have target space-times solving the usual supergravity equations of motion \cite{Sfetsos:2014cea,Demulder:2015lva,Borsato:2016zcf,Chervonyi:2016ajp}. There is also evidence that the latter class corresponds to a quantum group deformation of the gauge theory, but with $q$ a root of unity \cite{Hollowood:2015dpa}.

If instead the $r$-matrix solves the unmodified cYBE (a homogeneous $r$-matrix), first considered in \cite{Kawaguchi:2014qwa}, 
the YB $\sigma$-models have been demonstrated to give a wide variety of integrable target space-times including those generated by TsT transformations \cite{Matsumoto:2014nra,Matsumoto:2015uja,Matsumoto:2014gwa,Matsumoto:2015jja,vanTongeren:2015soa,Kyono:2016jqy,Osten:2016dvf}. For these models the corresponding dual theory can be understood in terms of a non-commutative $\mathcal{N} = 4$ super Yang-Mills with the non-commutativity governed by the $r$-matrix and the corresponding Drinfel'd twist \cite{vanTongeren:2015uha,vanTongeren:2016eeb}. Recently it has been shown that such YB $\sigma$-models can be also be understood in terms of non-abelian T-duality: given an $r$-matrix one can specify a (potentially non-abelian) group of isometries of the target space with respect to which one should T-dualise \cite{Hoare:2016wsk}. The deformation parameter appears by first centrally extending this isometry group and then T-dualising. Following a Buscher-type procedure, the Lagrange multiplier corresponding to the central extension is non-dynamical. In particular it is frozen to a constant value and thereby plays the role of the deformation parameter. This conjecture was proven in the NS sector in \cite{Borsato:2016pas}, where a slightly different perspective was also given. If one integrates out only the central extension, the procedure above can be seen to be equivalent to adding a total derivative $B$-field constructed from a 2-cocycle on the isometry group with respect to which we dualise and then dualising.

In this note we develop this line of reasoning. We begin by outlining the essential features of Yang-Baxter $\sigma$-models and the technology of non-abelian T-duality in Type II supergravity. After demonstrating that a centrally-extended T-duality can be reinterpreted as as non-abelian T-duality of a coset based on the Heisenberg algebra, we show how the machinery of non-abelian T-duality developed for Type II backgrounds can be readily applied to the construction of \cite{Hoare:2016wsk,Borsato:2016pas}. We confirm that the centrally-extended non-abelian T-duals produce the full Type II supergravity backgrounds corresponding to $\beta$-deformations (when the duality takes place in the $S^5$ factor of $AdS_5\times S^5$), non-commutative deformations (when performed in the Poincar\'e patch of $AdS_5$) and dipole deformations (when performed in both the $S^{5}$ and $AdS^{5}$ simultaneously). In appendices \ref{app:sugra} and \ref{app:algconv} we outline our conventions for supergravity and certain relevant algebras respectively. As a third appendix \ref{app:furtherexamples} we include some additional worked examples including one for which the non-abelian T-duality is anomalous and the target space solves the generalised supergravity equations.

The supergravity backgrounds in this note have appeared in the literature in the past but the derivation and technique presented here is both novel, simple and, we hope may have utility in the construction of more general supergravity backgrounds.

\section{Yang Baxter \texorpdfstring{$\sigma$}{sigma}-models}\label{sec:yangbaxter}

Given a semi-simple Lie algebra $\mathfrak{f}$ (and corresponding group $F$) we define an antisymmetric operator $R$ obeying
\begin{equation}\label{eq:cybe}
[R X , R Y] - R\left([R X, Y]+ [X,RY] \right) = c [ X, Y] \ , \quad X,Y \in \mathfrak{f} \ ,
\end{equation}
where the cases $c=\pm 1$ and $c=0$ are known as the classical and modified classical Yang Baxter equations (cYBE and mcYBE) respectively. We adopt some notation $X\wedge Y = X\otimes Y - Y \otimes X$ and define e.g.
\begin{equation}
r= T_1 \wedge T_2 + T_3 \wedge T_4 +\dots \ , \quad RX = \Tr_2 ( r (\mathbb{I}\otimes X)) \ .
\end{equation}

We define an inner product by the matrix trace of generators, $\Tr(T_{A} T_{B})$, and
lower and raise indices with this inner product and its inverse. In this way the $r$-matrix acts as
\begin{equation}\label{eq:rmatdef}
R(T_{A}) \equiv R_{A}{}^{B}T_{B} \ , \quad R_{A}{}^{B} = \Tr\left( \Tr_2 ( r (\mathbb{I}\otimes T_{A})) T^{B} \right) \ .
\end{equation}

Suppose we have a $\mathbb{Z}_{2}$ grading $\mathfrak{f} = \mathfrak{g}\oplus \mathfrak{k}$ for a subgroup $\mathfrak{g}$. Let $T_{A}$ be generators for $\mathfrak{f}$, $T_{\alpha}$ those of $\mathfrak{g}$ and $T_{i}$ the remaining orthogonal generators of $\mathfrak{k}$. We introduce a projector to the coset defined by $P(T_{\alpha})= 0$ and $P(T_{i})= T_{i}$ or, in matrix form,
\begin{equation}
P(T_{A}) \equiv P_{A}{}^{B}T_{B} \ , \quad P_{A}{}^{B} = \Tr\left( P( T_{A}) T^{B} \right) \ .
\end{equation}
We also define the adjoint action for $g\in F$ by
\begin{equation}
\Ad_{g} T_{A} \equiv gT_{A} g^{-1} \equiv D_{A}{}^{B}(g) T_{B} \ , \quad D_{AB} = \Tr(g T_{A} g^{-1} T_{B} ) \ .
\end{equation}

Let the two-dimensional worldsheet field $g$ be a coset representative for $F/G$ with which we define the currents
\begin{equation}
J_\pm = J_{\pm}^{A}T_{A} = g^{-1} \partial_{\pm} g \ , \quad J_{\pm}^{A}= \Tr(g^{-1} \partial_{\pm} g T^{A}) \ ,
\end{equation}
where we use light-cone coordinates $\partial_\pm = \partial_0 \pm \partial_1$.

The standard (bosonic) $\sigma$-model whose target is the coset space $F/G$ is
\begin{equation}\label{eq:cosetPCM}
{\cal L } = \Tr(J_+ P(J_-) ) \ .
\end{equation}
To define the Yang-Baxter model first we let
\begin{equation}
R_{g} = \Ad_{g^{-1}} R \Ad_{g} \ , \quad (R_{g})_{A}{}^{B} = D(g)_{A}{}^{C}R_{C}{}^{D}D(g^{-1})_{D}{}^{B} \ ,
\end{equation}
and define the operator
\begin{equation}
{\cal O} = \mathbb{I} - \eta R_{g}P \ , \quad {\cal O} _{A}{}^{B} = \delta_{A}{}^{B} - \eta P_{A}{}^{C} (R_{g})_{C}{}^{B} \ ,
\end{equation}
in which we have explicitly introduced the deformation parameter $\eta$. Later we will restrict to the case $c=0$ in \eqref{eq:cybe}, in which case the parameter $\eta$ can be absorbed into the definition of $R$.
The Yang-Baxter $\sigma$-model on a coset is given by \cite{Delduc:2014kha,Matsumoto:2014nra}
\begin{equation}
{\cal L } = \Tr(J_+ P( {\cal O}^{-1}J_-) ) = J_{+}^{A} E_{AB} J_{-}^{B} \ , \quad E_{AB} = {\cal O}^{-1}{}_{B}{}^{C} P_{CA} \ .
\end{equation}

\section{Non-abelian T-duality Technology}\label{sec:natd}

In this section we will mainly follow the approach of \cite{Sfetsos:2010uq,Lozano:2011kb,Itsios:2013wd} including the transformation of RR fluxes. Some subtleties are caused by the dualisation in a coset space and the approach here is slightly different to the one in \cite{Lozano:2011kb}.

Let us consider the standard (bosonic) $\sigma$-model whose target is the coset space $F/G$ whose Lagrangian is given in eq. \eqref{eq:cosetPCM},
and perform the non-abelian T-dual with respect to a subgroup $H \subset F$ (which need not, and in our applications mostly will not be, either semi-simple or a subgroup of $G$). Let $H_a$ be the generators of $\mathfrak{h}$ and $\tilde{H}^a$ generators of a dual algebra $\mathfrak{h}^\star$ normalised such that $\Tr(H_a \tilde{H}^b)= \delta^b_a$.

Let us we parametrise the coset representative as $g= h \hat{g}$. We define $\hat{J}= \hat{g}^{-1} d \hat{g}$ and $L = L^a H_a = h^{-1}dh$ such that
\begin{equation}
J= \hat{J}+ L^a H_a^{\hat{g} }\ , \quad H_a^{\hat{g} } =\Ad_{\hat{g}^{-1}} H_a \ .
\end{equation}
We also define
\begin{equation}
\begin{aligned}
G_{ab} &= \Tr( H^{\hat{g}}_{a} P( H^{\hat{g}}_{b} ) ) \ , \quad
Q_{a} &= \Tr( \hat{J} P( H^{\hat{g}}_{a} ) ) \ .
\end{aligned}
\end{equation}
In this notation the $H$ isometry of the target space is manifest since the metric corresponding to eq. \eqref{eq:cosetPCM} is
\begin{equation}
ds^2 = \Tr( \hat{J} P(\hat{J} )) + 2 Q^T L + L^T G L = \Tr( \hat{J} P(\hat{J} )) - Q^TG^{-1} Q + e^T e \ ,
\end{equation}
where we introduce the frame fields
\begin{equation}
G= \kappa^T \kappa \ , \quad e= \kappa \left( L + G^{-1} Q \right) .
\end{equation}

We perform the dualisation by introducing a $\mathfrak{h}$-valued connection with components $A_{\pm} = A_{\pm}^{a}H_{a}$ and a $\mathfrak{h}^\star$-valued Lagrange multiplier $V= v_{a} \tilde{H}^{a}$. We covariantise currents
\begin{equation}
J^{\nabla}_{\pm} = g^{-1} d g + g^{-1} A_{\pm } g \ ,
\end{equation}
such that we are gauging a left action of some $\tilde{h} \in H$
\begin{equation}
g \rightarrow \tilde{h} g \ , \quad A \rightarrow \tilde{h} A \tilde{h} ^{-1} - d \tilde{h} \tilde{h}^{-1} \ ,
\end{equation}
and consider
\begin{equation}
{\cal L }^{ \nabla} = \Tr(J^{\nabla}_+ P(J^{\nabla}_-) ) + \Tr(V F_{+ -} ) \ ,
\end{equation}
where the field strength is $F_{+-} = \partial_{+} A_{-} - \partial_{-} A_{+} + [A_{+} , A_{-}]$.

We continue by gauge fixing on the group element $g = \hat{g}$ i.e. $h=1$.\footnote{In some cases it can be that this doesn't fully fix the gauge and additional fixing should be imposed on the Lagrange multipliers $ V= v_a \tilde{H}^a$, details of this are discussed in \cite{Lozano:2011kb}.} Integrating the Lagrange multipliers enforces a flat connection and one recovers the starting model since
\begin{equation} \label{eq:puregauge}
A_\pm = h^{-1}\partial_\pm h = L_\pm \ ,
\end{equation}
and upon substituting back into the action one recovers the starting $\sigma$-model.

On the other hand, integrating by parts the derivative terms of the gauge fields yields
\begin{equation}
{\cal L }^{ \nabla} = \Tr(\hat{J}_{+} P(\hat{J}_-) ) + A_{+}^{a}A_{-}^{b} M_{ab} + A_{+}^{a}( \partial_{-} v_{a}+ Q_{-a} ) - A_{-}^{a}( \partial_{+} v_{a} - Q_{+a} ) \ ,
\end{equation}
in which we have pulled back the one-forms $Q$ and $\hat{J}$ to the worldsheet and defined
\begin{equation}
\begin{aligned}
F_{ab} &= \Tr([H_{a} ,H_{b}]V) = f_{ab}{}^{c} v_{c} \ , \quad M_{ab} =G_{ab} + F_{ab} \ .
\end{aligned}
\end{equation}
The gauge field equations of motion now read
\begin{equation}\label{eq:gauge}
A_{-} = - M^{-1} ( \partial_{-} v + Q_{-} ) \ , \quad A_{+} = M^{-T} ( \partial_{+} v - Q_{+ } ) \ .
\end{equation}
Combining these equations of motion for the gauge field in eqs.~\eqref{eq:gauge} and \eqref{eq:puregauge} sets up the canonical transformation between T-dual theories. Substitution of the gauge field equation of motion \eqref{eq:gauge} into the action yields the T-dual model given by
\begin{equation}\begin{split}\label{eq:dualmodellag}
{\cal L }_{dual} & = \Tr(\hat J_+ P (\hat J_-)) - A_+^T M A_-
\\ & = \Tr(\hat{J}_{+} P(\hat{J}_-) ) + ( \partial_{+} v_{a} - Q_{+a} )(M^{-1})^{ab} ( \partial_{-} v_{b} + Q_{-b} ) \ ,
\end{split}\end{equation}
where in the first line $A_\pm$ are evaluated on the gauge field equation of motion eq. \eqref{eq:gauge}.

The NS fields can be read directly from this $\sigma$-model and in particular the dual metric is given as
\begin{equation}
\widehat{ds}^2 = \Tr( \hat{J} P(\hat{J} )) - Q^TG^{-1} Q + \widehat{e}_\pm^T \widehat{e}^{\vphantom{T}}_\pm \ ,
\end{equation}
with $\widehat{e}_\pm$ given by the push forwards to target space of
\begin{equation}\label{eq:dualframe}
\widehat{e}_\pm = \kappa \left( A_\pm + G^{-1} Q_\pm \right) \ ,
\end{equation}
evaluated on the gauge field equation of motion \eqref{eq:gauge}. On the worldsheet left and right moving fermionic sectors couple to the frame fields $\widehat{e}_{+}$ and $\widehat{e}_-$ respectively. Since they define the same metric they are related by a local Lorentz rotation
\begin{equation}\label{eq:Lorentztrans}
\Lambda \widehat{e}_{-} = \widehat{e}_{+} \ , \quad \Lambda = -\kappa M^{-T}M \kappa^{-1}
\end{equation}
This Lorentz rotation lifts to spinors via
\begin{equation}\label{eq:Spinortrans}
\Omega^{-1}\Gamma^a \Omega= ( \Lambda\cdot\Gamma)^a \ .
\end{equation}
Using the Clifford isomorphism we convert the poly-form sum of RR fluxes
\begin{equation}\label{eq:Polyform}
{\cal P}= e^{\Phi} ( F_{1 } + F_{3} + F_{5} - \star F_3 + \star F_1 ) \ ,
\end{equation}
to a bi-spinor matrix. The T-duality rule is then given by
\begin{equation}
\widehat{{\cal P}} ={\cal P} \cdot \Omega^{-1} \ .
\end{equation}
The relationship between the local Lorentz rotations and RR field transformation in the case of abelian T-duality in curved space was made explicit in the work of Hassan \cite{Hassan:1999bv} and developed in the present context in \cite{Sfetsos:2010uq}. Note that although we have "bootstrapped" the transformation rule for the RR sector from knowledge of the NS sector it seems rather likely that the same conclusion can be reached in e.g. the pure spinor superstring by a straightforward extension of the arguments presented for abelian \cite{Benichou:2008it} and fermionic T-duality \cite{Sfetsos:2010xa}.\footnote{An explicit demonstration of the RR transformation law in the context of supersymmetry in $SU(2)$ non-Abelian T-duality can be found in\cite{Kelekci:2014ima}.}

\

Finally let us turn to the transformation of the dilaton field under non-abelian T-duality. For the non-abelian duality to preserve conformality the dualisation procedure must avoid the introduction of a mixed gravitational-gauge anomaly \cite{Alvarez:1994np,Elitzur:1994ri} and the structure constants of the algebra in which we dualise should satisfy 
\begin{equation}
n_a \equiv f_{ab}{}^b = 0 \ .
\end{equation}
When this is the case the dual dilaton comes from the Gaussian integration in the path integral \cite{Buscher:1987qj}
\begin{equation}\label{eq:diltrans}
\widehat{\Phi} = \Phi - \frac{1}{2}\log \det M \ .
\end{equation}

On the other hand if $n_a \neq 0$ the dual model is not expected to be conformal, however it will be globally scale invariant. In this case we still define the dual ``dilaton'' to be given by \eqref{eq:diltrans}. The global scale invariance then implies that, for example, the one-loop metric and $B$-field beta-functions (defined in \eqref{eq:betafunctions} of appendix \ref{app:sugra}) only vanish up to diffeomorphisms and gauge transformations. This is in contrast to the conformal case, for which the beta-functions of the metric, $B$-field and dilaton vanish identically, while the RR fluxes solve the first order equations in eq. \eqref{eq:sugraeq} of appendix \ref{app:sugra}.

It transpires that the globally scale invariant models that arise from dualising with $n_a \neq 0$ satisfy a stronger set of equations than those of global scale invariance \cite{Hoare:2016wsk}. These are a modification of the Type II supergravity equations \cite{Arutyunov:2015mqj,Wulff:2016tju,Sakatani:2016fvh} that depend on a particular Killing vector $I$ of the background such that when $I = 0$ standard Type II supergravity is recovered. These equations are given in eqs. \eqref{eq:modsugra1} of appendix \ref{app:sugra}.

As mentioned above we take the dual ``dilaton'' field in these equations to still be defined in terms of the original dilaton via the transformation \eqref{eq:diltrans}, while the one-forms $X$, $Z$ and $W$ are defined in terms of $\Phi$ and the Killing vector $I$ as in eq. \eqref{eq:modsugra3} of appendix \ref{app:sugra}.

To show that the dual background solves the modified supergravity equations we follow the derivation in \cite{Hoare:2016wsk}. After splitting the Lagrange multiplier as $v_a = u_a + y n_a$, it transpires that shifting $y$ is a symmetry of the dual background and T-dualising $y \to \tilde y$ gives a conformal $\sigma$-model with a dilaton linear in $\tilde y$. From the results of \cite{Arutyunov:2015mqj} this then implies that, in our conventions, the dual model solves the modified supergravity equations with $I^y = - 1$.

The classical bosonic string Lagrangian in conformal gauge,
\begin{equation}\label{eq:cbsacg}
\mathcal{L} = \partial_+ x^m (G_{mn} + B_{mn}) \partial_- x^n \ ,
\end{equation}
has the property that when we replace $\partial_- x^m \to I^m$ it equals $W_n \partial_+ x^n$ where the one-form $W$, defined in eq. \eqref{eq:modsugra3}, is given by
\begin{equation}
W_n = I^m (G_{mn} - B_{mn}) \ .
\end{equation}
Following this procedure in the dual model \eqref{eq:dualmodellag} with $I^y = - 1$ and the remaining components vanishing, we find that
\begin{equation}\label{eq:Ampush}
W_n \partial_+ x^n = - A_+^a n_a \ ,
\end{equation}
with $A_+$ evaluated on the gauge field equation of motion \eqref{eq:gauge}. To summarise; if the T-duality is anomalous then the background solves the modified supergravity equations with the one-form $W$, which can be used to define the modification, given by the push forward of the $A_+$ component of the gauge field evaluated on its equations of motion.

\section{Centrally-extended duality}\label{sec:centralext}

Let us now consider non-abelian T-dualities with respect to centrally-extended algebras. In particular we consider the setup considered in \cite{Hoare:2016wsk,Borsato:2016pas} in which case the dualities are equivalent to Yang-Baxter deformations for homogeneous $r$-matrices. The aim of this section is to extend this to the RR fluxes using the technology outlined in section \ref{sec:natd}. We start by recalling that for a homogeneous $r$-matrix for a Lie algebra $\mathfrak{f}$
\begin{equation}\label{eq:rmatans}
r= \sum_{j} \eta_j \, \big( \sum_{i=1}^{n(j)} a_{ij} \, X_{ij} \wedge Y_{ij} \big) \ ,
\end{equation}
the generators $\{X_{ij},Y_{ij}\}$ (for each fixed $j$) form a basis for a subalgebra $\mathfrak{h}$, which admits a central extension. In eq. \eqref{eq:rmatans} $\eta_{j}$ are free parameters, while $a_{ij}$ are fixed real coefficients. For each free parameter we introduce a central extension, such that the centrally-extended algebra has a basis $\mathfrak{h}^{ext} = \{X_{ij}, Y_{ij}\} \oplus \{Z_j\}$, with commutation relations $[X_{ij},Y_{ij}]^{ext} = [X_{ij},Y_{ij}] + a_{ij}^{-1} Z_j$ (for fixed $i$ and $j$), and $[X_{ij},Z_j]^{ext} = [Y_{ij},Z_j]^{ext} = 0$. This is the centrally-extended algebra with respect to which we dualise.

The precise relation between the centrally-extended non-abelian T-dual and the Yang-Baxter deformation was made in the NS sector in \cite{Borsato:2016pas}.
The $R$-operator (see eq. \eqref{eq:rmatdef}) governing a certain Yang-Baxter deformation defines an invertible map from $\mathfrak{h}^\star$ to $\mathfrak{h}$. Recalling our parametrisation of the $F/G$ coset representative $g= h \hat{g}$ with $h \in H$, we may write $h = \exp(R(X))$ for $X\in \mathfrak{h}^\star$. If $\mathfrak{h}$ is abelian then the relation between the Lagrange multipliers parametrising the T-dual model and the YB deformed model is simple: $V= \eta^{-1} R(X)$. When $\mathfrak{h}$ is non-abelian the relation is more involved \cite{Borsato:2016pas}.

\

One can formally set up the non-abelian T-dual of the central extension by considering the coset of the centrally-extended algebra by the central generators. To see this precisely let us consider the Heisenberg algebra, i.e. the central extension of $U(1)^2$
\begin{equation}
[X, Y ]= Z \ , \quad [X , Z] = [ Y,Z ] = 0 \ .
\end{equation}
We let $T_{1}= X, T_{2}= Y$ and $T_{3} = Z$ and hence the only non-vanishing structure constant is $f_{12}{}^{3}=1$.
We introduce the matrix generators
\begin{equation}
T_{1} = \left( \begin{array}{ccc}
0 & 1 & 0 \\
0 & 0 & 0 \\
0 & 0 & 0 \\
\end{array} \right) \ , \quad T_{2} = \left( \begin{array}{ccc}
0 & 0 & 0 \\
0 & 0 & 1 \\
0 & 0 & 0 \\
\end{array} \right) \ , \quad T_{3 }= \left( \begin{array}{ccc}
0 & 0 & 1 \\
0 & 0 & 0 \\
0 & 0 & 0 \\
\end{array} \right) \ ,
\end{equation}
and the group element
\begin{equation}
g= \exp \left[ x_{1} T_{1} + x_{2}T_{2} + (x_{3}-\frac{1}{2} x_{1}x_{2} ) T_{3}\right] = \left( \begin{array}{ccc}
1& x_{1} &x_{3} \\
0 &1 & x_{2}\\
0 & 0 & 1 \\
\end{array} \right) \ .
\end{equation}
The left-invariant one-forms $g^{-1}dg= L^{i} T_{i}$ are
\begin{equation}
L^{1} = dx_{1} \ , \quad L^{2 } = dx_{2} \ , \quad L^{3} = dx_{3} - x_{1} dx_{2} \ .
\end{equation}

We consider a $\sigma$-model based on this algebra
\begin{equation}\label{eq:Heisenberg}
{\cal L} = E_{ab} L_+^a L_-^b = f_1 L^{1}_+ L^{1}_- + f_2 L^{2}_+ L^{2}_- + \lambda L^{3}_+ L^{3}_- \ ,
\end{equation}
i.e. $E = \operatorname{diag} (f_1,f_2, \lambda)$, where we allow $f_{1,2}$ to be functions of any spectator coordinates. In the limit $\lambda \rightarrow 0$ the theory develops a gauge invariance (the coordinate $x_3$ drops out of the action all together) and reduces to the $\sigma$-model whose target space is simply $ds^2 = f_1 dx_1^2 + f_2 dx_2^2$. This Rube Goldberg construction allows us to now go head and perform a non-abelian T-duality on the coset following the techniques of \cite{Lozano:2011kb}.

The resulting dual $\sigma$-model is given by
\begin{equation}
\mathcal{L}_{dual} = \partial_+ v_a (M^{-1})^{ab} \partial_- v^b
\end{equation}
in which
\begin{equation}\begin{split}
M_{ab} = E_{ab} + f_{ab}{}^c v_c & = \left( \begin{array}{ccc}
f_1 & v_3 & 0 \\
-v_3 & f_2 & 0 \\
0 & 0 & \lambda \end{array} \right) \ ,
\\
(M^{-1})^{ab} & = \left( \begin{array}{ccc}
h f_2 & - h v_3 & 0 \\
h v_3 & h f_1 & 0 \\
0 & 0 & \frac{1}{\lambda} \end{array} \right)
\ , \quad h= \frac{1}{f_1 f_2 + v_3^2}\ .
\end{split}\end{equation}
The matrix $M^{-1}$ diverges in the limit of interest $\lambda \to 0$. In particular, the coefficient of the kinetic term for $v_3$ becomes infinite in the limit and this can be understood as freezing $v_3$ to a constant value. To see this let us rewrite the dual $\sigma$-model as
\begin{equation}
\mathcal{L}_{dual} = \partial_+ v_\alpha (M^{-1})^{\alpha\beta} \partial_- v^\beta + \lambda a_+ a_- + a_+ \partial_-v_3 - a_- \partial_+ v_3 \ , \quad \alpha,\beta = 1,2 \ ,
\end{equation}
where we integrate over $a_\pm$. Now taking $\lambda \to 0$ and then integrating out $a_\pm$ we find $\partial_\pm v_3 = 0$ and indeed $v_3$ is frozen to a constant value. This final step is analogous to the Buscher procedure considered in \cite{Hoare:2016wsk}. The true target space of the dual model is then spanned by the coordinates $v_1 \equiv y_2$ and $v_2 \equiv y_1$, while $v_3 \equiv \nu$ is a constant parameter. The dual metric, B-field and dilaton shift are easily ascertained:
\begin{equation}\label{eq:eq1}
\widehat{ds}^2 = h ( f_1 dy_1^2 + f_2 dy_2^2 ) \ , \quad \widehat{B} = \nu h dy_1 \wedge dy_2 \ , \quad \widehat{\Phi} = \Phi + \frac{1}{2}\log h \ .
\end{equation}
Frame fields for the dual geometry as seen by left and right movers \cite{Lozano:2011kb} are given by
\begin{equation}\label{eq:eq2}
\widehat{e}_{+}^{\, i} = (\kappa\cdot M^{-1})^{a i} d v_a \, \quad \widehat{e}_{-}^{\,i} = - (\kappa \cdot M^{-1})^{ i a } dv_a \ , \quad i=1,2 \ , \quad a = 1,2,3 \ .
\end{equation}
where $\frac{1}{2} (E+ E^T)= \kappa^T \kappa$. Explicitly we have
\begin{equation}\label{eq:eq3}
\widehat{e}_+ = \left(\begin{array}{c} h \sqrt{f_1} (f_2 dy_2 + \nu dy_1) \\ h \sqrt{f_2}(f_1 dy_1 - \nu d y_2) \end{array}\right) \ , \quad
\widehat{e}_- = \left(\begin{array}{c} h \sqrt{f_1} (-f_2 dy_2 + \nu dy_1) \\ h \sqrt{f_2}( -f_1 dy_1 - \nu d y_2) \end{array}\right) \ .
\end{equation}
The plus and minus frames are then related by a Lorentz rotation
\begin{equation}\label{eq:eq4}
\Lambda \cdot \widehat{e}_- = \widehat{e}_+ \ , \quad \Lambda = h \left(\begin{array}{cc}
\nu^2 -f_1f_2 & - 2 \nu \sqrt{f_1 f_2} \\
2\nu\sqrt{f_1 f_2} & \nu^2 - f_1 f_2 \end{array}\right) \ , \quad \det \Lambda = 1 \ , \quad \Lambda \cdot \Lambda^T = \mathbb{I} \ .
\end{equation}

This coset-based construction is interesting, however for calculation purposes it is enough to follow the T-duality rules for the non-centrally-extended dualisation, while replacing the structure constants entering the $\dim H \times \dim H$ matrix $F_{ab}= \Tr([ H_{a}, H_{b}] V)$ with the corresponding central extension and the centrally-extended Lagrange multipliers i.e. $V^{ext} = v_{a} H^{a} + v_{\mu }Z^{\mu}$ and $F^{ext}_{ab}= \Tr([ H_{a}, H_{b}]^{ext} V^{ext})$.

\section{Applications}\label{sec:examples}

Let us now turn to specific examples for which we construct the dual RR fluxes corresponding to various centrally-extended non-abelian T-dualities of $AdS_5 \times S^5$ using the technology outlined in section \ref{sec:natd}. Here we will consider certain deformations that are well-known to correspond to TsT transformations. In appendix \ref{app:furtherexamples} we consider further examples that correspond to Yang-Baxter deformations with time-like abelian and non-abelian $r$-matrices.

\subsection{Application 1: Non-Commutative Deformations}\label{ssec:app1}
The first application we consider is the string background dual to non-commutative $\mathcal{N} = 4$ super Yang-Mills \cite{Hashimoto:1999ut,Maldacena:1999mh}
\begin{align}\nonumber
ds^2 &= \frac{du^2}{u^2} + u^2 \left( -dt^2 + dx_1^2 + \tilde h (dx^2_2 + dx^2_3) \right) + d\Omega_5^2 \ ,
\quad
\tilde h = \frac{1}{1+ a^4 u^4} \ ,
\\ \label{eq:mrback}
B &= a^2 \tilde h u^4 dx_2 \wedge dx_3 \ , \quad \exp 2 \Phi = g_0^2 \tilde h \ , \\ \nonumber
F_3 &= -\frac{4}{g_0} a^2 u^3 dt\wedge dx_1 \wedge du \ , \quad
F_5= \frac{4}{g_0} \tilde h u^3 (1+\star) \, du \wedge dt \wedge dx_1 \wedge dx_2 \wedge dx_3 \ .
\end{align}

Starting from the undeformed background
\begin{equation}\label{eq:undefads5}
\begin{aligned}
ds^2 &= \frac{du^2}{u^2} + u^2 \left( -dt^2 + dx_1^2 + dx^2_2 + dx^2_3 \right) + d\Omega_5^2 \ ,
\quad \exp 2 \Phi = g_0^2 \ ,
\\
F_5& = \frac{4}{g_0} u^3 (1+\star) \, du \wedge dt \wedge dx_1 \wedge dx_2 \wedge dx_3 \ ,
\end{aligned}
\end{equation}
we now consider the non-abelian T-dual with respect to the central extension of $U(1)^2$, where the $U(1)^2$ is generated by shifts in $x_2$ and $x_3$. Using eqs. \eqref{eq:eq1} -- \eqref{eq:eq4} with $y_1 = \frac{x_3}{a^2}$, $y_2 = \frac{x_2}{a^2}$, $f_1 = f_2 = u^2$ and setting the deformation parameter $\nu=a^{-2}$ we find that the plus and minus frames are given by
\begin{equation}
\widehat e_+ = \left(\begin{array}{c} \frac{hu}{a^4} ( a^2 u^2 dx_2 + dx_3) \\ \frac{hu}{a^4}( a^2 u^2 dx_3 - dx_2) \end{array}\right) \ , \quad
\widehat e_- = \left(\begin{array}{c} \frac{hu}{a^4} (- a^2 u^2 dx_2 + dx_3) \\ \frac{hu}{a^4}(- a^2 u^2 dx_3 - dx_2) \end{array}\right) \ , \quad
h = \frac{a^4}{1+a^4 u^4} \ .
\end{equation}

The Lorentz rotation of \eqref{eq:eq4} induces a spinorial action according to \eqref{eq:Spinortrans} given by
\begin{equation}
\Omega = \sqrt{\frac{h}{a^4}} \left( \mathbb{I} - a^2 u^2 \Gamma^{23} \right) \ .
\end{equation}
Now let us consider the duality transformation of the five-form RR flux supporting the $AdS_5 \times S^5$ geometry \eqref{eq:undefads5}. The self-dual five-form flux can be written as $F_5 = (1+\star) f_5$, where
\begin{equation}
f_5 = \frac{4}{g_0}u^3 du \wedge dt \wedge d x_1\wedge dx_2 \wedge dx_3 \equiv \frac{4}{g_0} e^u \wedge e^0 \wedge e^1 \wedge e^2 \wedge e^3 \ .
\end{equation}
The corresponding poly-form of eq.~\eqref{eq:Polyform} is then given by
\begin{equation} {\cal P} = 4 \Gamma^{u 0 1 2 3} - 4 \Gamma^{56789} \ .
\end{equation}
The transformation of the poly-form under T-duality is given by
\begin{equation}
\widehat{{\cal P}}= {\cal P} \cdot \Omega^{-1} = 4\sqrt{\frac{h}{a^4}} \Gamma^{u 0 1 2 3} - 4\sqrt{\frac{h}{a^4}}a^2 u^2 \Gamma^{u 0 1 } + \text{duals} \ .
\end{equation}
Extracting the dual background from the above data we find
\begin{equation}\label{eq:mrback2}
\begin{aligned}
\widehat{ds}^2 &= \frac{du}{u^2} + u^2 \big( -dt^2 + dx_1^2 + \frac{h}{a^4} ( dx_2^2 + dx_3^2 ) \big) + d\Omega_5^2 \ , \\
\widehat{B}&= -a^{-2}\frac{h}{a^4} dx_2 \wedge dx_3 \ , \quad
\exp(2\widehat{\Phi}) = (g_0a^2)^2 \frac{h}{a^4} \ , \\
\widehat{F}_3&= -\frac{4}{g_0 a^2} a^2 u^3 du \wedge dt \wedge dx_1 \ , \quad
\widehat{F}_5 = \frac{4}{g_0 a^2} \frac{h}{a^4} u^3 (1+\star)\, du \wedge dt \wedge dx_1 \wedge dx_2 \wedge dx_3 \ . \\
\end{aligned}
\end{equation}
Noting that $\tilde h = a^{-4}h$, we then immediately see that this is precisely the background \eqref{eq:mrback} up to the constant shift of the dilaton $g_0 \to g_0 a^{-2}$.
A small subtlety is that while there is precise agreement between $H = dB$ in \eqref{eq:mrback} and \eqref{eq:mrback2}, the $B$-field itself differs by a gauge:
\begin{equation}
\widehat{B} =- a^{-2} \frac{1}{1+a^4 u^4} dx_2 \wedge dx_3 = -a^{-2}dx_2 \wedge dx_3 + \frac{ a^2 u^4}{1+a^4 u^4}dx_2 \wedge dx_3 \ .
\end{equation}
This is always the case in these comparisons \cite{Hoare:2016wsk,Borsato:2016pas} and from now on by agreement we always mean up to a gauge term in the $B$-field.

\subsection{Application 2: Marginal Deformations}\label{ssec:app2}

${\cal N}=4$ super Yang-Mills with gauge group $SU(N)$ admits a class of marginal deformations that preserve ${\cal N}= 1$ supersymmetry \cite{Leigh:1995ep}. The corresponding superpotential for these theories is
\begin{equation}
W = \kappa \Tr \Big( \Phi_1 [\Phi_2, \Phi_3]_q + \frac{h}{4}\big( \sum_{i=1}^3 \Phi_i^2 \big) \Big) \ ,
\end{equation}
in which the commutator is $q$-deformed i.e. $[\Phi_i, \Phi_j]_{q} = \Phi_i \Phi_j - q \Phi_j \Phi_i$. For the case where $h=0$ and $q= e^{i \beta}$ with $\beta$ real, known as the $\beta$-deformation, the seminal work of Lunin and Maldacena \cite{Lunin:2005jy} provides the gravitational dual background constructed via a TsT solution generating technique consisting of a sequence of T-duality, coordinate shift and T-duality. In this case integrability has been shown on both the string \cite{Frolov:2005ty,Frolov:2005dj,Alday:2005ww} and gauge side \cite{Roiban:2003dw,Berenstein:2004ys,Frolov:2005ty,Beisert:2005if} of the AdS/CFT correspondence. The cubic deformation ($q=1$ and $h \neq 0$) is far less understood, with integrability not expected and, as of now, no known complete gravitational dual constructed.

A more general class of non-supersymmetric deformations\footnote{Care needs be taken in the interpretation of this deformation. Away from the supersymmetric point the $\gamma_i$ deformation is not conformal due a running coupling of a double-trace operator \cite{Fokken:2013aea} and indeed the gravitational dual has a tachyon \cite{Spradlin:2005sv}.} of this gauge theory are defined by a scalar potential
\begin{equation}
V= \Tr \Big( |[\Phi_1, \Phi_2]_{q_3}|^2 + |[\Phi_2, \Phi_3]_{q_1}|^2 + |[\Phi_3, \Phi_1]_{q_2}|^2 \Big) + \Tr \Big( \sum_{i=1}^3 [\Phi_i , \bar{\Phi}_i] \Big)^2 \ ,
\end{equation}
where $q_i = e^{-2\pi i \gamma_i}$. This three parameter deformation, known as the $\gamma$-deformation, enjoys integrability both in the gauge theory \cite{Frolov:2005iq} and in the worldsheet $\sigma$-model with the target space given by the postulated gravitational dual background constructed in \cite{Frolov:2005dj}. Upon setting all three deformation parameters equal this reduces to the $\beta$-deformation with enhanced ${\cal N}=1$ supersymmetry and hence we will proceed with the general case.

Rather remarkably the string $\sigma$-model in the $\gamma$-deformed target space can be obtained as Yang-Baxter $\sigma$-model \cite{Kyono:2016jqy,Osten:2016dvf}. Let us consider the bosonic sector, restricting our attention to the five-sphere of $AdS_5\times S^5$; the $AdS$ factor plays no role in what follows. It is convenient to follow \cite{Frolov:2005dj} and parametrise the $S^5$ in coordinates adapted to the $U(1)^3$ isometry
\begin{equation}\label{eq:metgam}
ds_{S^5}^2 = d\alpha^2 + \S_\alpha^2 d\xi^2 + \C_\alpha^2 d\phi_1^2 + \S_\alpha^2 \C_\xi^2 d\phi_2^2 + \S_\alpha^2 \S_\xi^2 d\phi_3^2
= \sum_{i= 1\dots 3} dr_i^2 + r_i^2 d\phi_i^2 \ ,
\end{equation}
where $r_1 = \C_\alpha$, $r_2= \S_\alpha \C_\xi$, $r_3= \S_\alpha \S_\xi$ with $\C_x$ and $\S_x$ denoting $\cos x$ and $\sin x$ respectively. The sphere can be realised as the coset $SU(4)/SO(5)$ for which a particular coset representative is given by
\begin{equation}\label{eq:paragam}
g = e^{\frac{1}{2} \sum_{m=1}^3 \phi^m h_m } e^{-\frac{\xi}{2} \gamma^{13}} e^{\frac{i}{2} \alpha \gamma^1} \ ,
\end{equation}
where $ \gamma^{13}$ and $\gamma^1$ are certain $SU(4)$ generators (see appendix \ref{app:algconv} for conventions) and $h_i$ are the three Cartan generators. Letting $P$ be the projector onto the coset and $J_\pm = g^{-1} \partial_\pm g$ pull backs of the left-invariant one-form, the $S^5$ $\sigma$-model Lagrangian is
\begin{equation}
{\cal L } = \Tr(J_+ P(J_-) ) \ ,
\end{equation}
with the parametrisation \eqref{eq:paragam} giving the $\sigma$-model with target space metric \eqref{eq:metgam}.

Starting with the $r$-matrix
\begin{equation}
r= \frac{\nu_1}{4} h_2 \wedge h_3 + \frac{\nu_3}{4} h_1 \wedge h_2 + \frac{\nu_2}{4} h_3 \wedge h_1 \ ,
\end{equation}
it was shown in \cite{Matsumoto:2014nra,vanTongeren:2015soa} that the NS sector of the Yang-Baxter $\sigma$-model matches the $\gamma$-deformed target space explicitly given by
\begin{equation}\label{eq:gammadef}
\begin{aligned}
ds^2 & = ds^2_{AdS}+ \sum_{i= 1\dots 3} ( dr_i^2 + G r_i^2 d\phi_i^2) + G r_1^2 r_2^2 r_3^2 \Big( \sum_{i= 1\dots 3} \nu_i dr_i \Big)^2 \ ,\\
B & = G ( r_1^2 r_2^2 \nu_3 d\phi_1 \wedge d\phi_2 + r_1^2 r_3^2 \nu_2 d\phi_3 \wedge d\phi_1 + r_2^2 r_3^2 \nu_1 d\phi_2 \wedge d\phi_3 ) \ , \\
\end{aligned}
\end{equation}
with
\begin{equation}
G^{-1}\equiv \lambda= 1+ r_1^2 r_2^2 \nu_3^2 + r_3^2 r_1^2 \nu_2^2 + r_2^2 r_3^2 \nu_1^2 \ ,
\end{equation}
where the parameters $\nu_i$ are related to the $\gamma_i$ of the field theory by a factor of the $AdS$ radius \cite{Frolov:2005dj}, which we suppress throughout.

We would like to interpret this in terms of the centrally-extended (non-)abelian T-duality introduced in section \ref{sec:centralext}. To do so we find it expedient to make a basis transformation of the Cartan generators; let us assume $\nu_3 \neq 0$ and define
\begin{equation}
\tilde{h}_1 = h_1 - \frac{\nu_1}{\nu_3} h_3 \ , \quad \tilde{h}_2 = h_2 - \frac{\nu_2}{\nu_3} h_3 \ , \quad \tilde{h}_3 = h_3+ \frac{\nu_1}{\nu_3} h_3 + \frac{\nu_2}{\nu_3} h_3 \ .
\end{equation}
In this basis the $r$-matrix simply reads
\begin{equation}
r= \frac{\nu_3}{4} \tilde{h}_1 \wedge \tilde{h}_2 \ .
\end{equation}
We also introduce a new set of angles such that $\tilde{h}_i \tilde{\phi}_i = h_i \phi_i$ (where the sum over $i$ is implicit). Written in this way it is clear that we should consider a centrally-extended (non-)abelian T-duality along the $\tilde{h}_1$ and $\tilde{h}_2$ directions. To proceed we defined a slightly exotic set of frame fields for the $S^5$, adapted to the dualisation as described
\begin{equation}
\begin{aligned}
e^\alpha &= d \alpha \ , \quad e^\xi = \sin \alpha d\xi \ , \quad
e^1 = \frac{1}{\varphi \sqrt{\lambda-1} } \left( r_1^2 \varphi^2 d\phi_1 - r_2^2 r_3^2 \nu_1 \nu_2 d\phi_2 - r_2^2 r_3^2 \nu_1 \nu_3 d\phi_3 \right) \ ,
\\
e^2 & = \frac{1}{\varphi } \left( r_2^2 \nu_3 d\phi_2 - r_3^2 \nu_2 d\phi_3 \right) \ , \quad
e^3 = \frac{r_1 r_2 r_3}{\sqrt{\lambda -1} }\sum_{i} \nu_i d\phi_i \ ,
\end{aligned}
\end{equation}
where $\varphi=(r_2^2 \nu_3^2 + r_3^3 \nu_2^2)^{\frac{1}{2}}$. Though these frames depend on $\nu_i$ the overall metric remains the round $S^5$ independent of $\nu_i$. The advantage of this basis is that the T-dualisation acts only on the $e_1$ and $e_2$ directions. We non-abelian T-dualise with respect to the central extension of $\tilde{h}_1$ and $\tilde{h}_2$ making the gauge fixing choice
\begin{equation}
\hat{g} = e^{\frac{1}{2} \tilde{\phi}_3 \tilde{h}_3 } e^{-\frac{\xi}{2} \gamma^{13}} e^{\frac{i}{2} \alpha \gamma^1}
\end{equation}
and by parametrising the Lagrange multiplier parameters as
\begin{equation}
v_1 = - \frac{2 }{\nu_{3}} \tilde{\phi}_2 \ , \quad v_2 =\frac{2 }{\nu_{3}} \tilde{\phi}_1 \ , \quad v_3 = \frac{4}{\nu_3} \ , \quad dv_3= 0 \ .
\end{equation}

After some work one finds the dual metric is exactly that of eq.~\eqref{eq:gammadef} with a $B$-field matching up to a gauge transformation.\footnote{As with the previous example the $B$-field obtained by the central extension dualisation procedure differs by a closed piece $\Delta B = \frac{1}{\nu_1^2+ \nu_2^2 + \nu_3^2} \left( \nu_1 d\phi_2\wedge d\phi_3 + \nu_3 d\phi_1\wedge d\phi_2+ \nu_2 d\phi_3\wedge d\phi_1 \right)$.\label{foot:bdiff}} 
 The dual dilaton is given by
\begin{equation}
e^{\widehat{\Phi} - \phi_0} = \frac{\nu_3}{4 \sqrt{\lambda}} \ .
\end{equation}

The frame fields produced by dualisation, using eq.~\eqref{eq:dualframe}, are
\begin{equation}
\begin{aligned}
\widehat{e}^{\,\alpha} &= e^\alpha \ , \quad \widehat{e}^{\,\xi} = e^\xi \ , \quad \widehat{e}^{\,3} = e^3 \ , \\
\widehat{e}^{\,1} &\equiv \widehat{e}^{\,1}_{+} = \frac{1}{ \lambda \varphi \sqrt{\lambda-1} } \left( r_1^2 \varphi^2 d\phi_1 - r_2^2 (r_3^2 \nu_1 \nu_2 + (\lambda-1)\nu_3 ) d\phi_2 - r_3^2 (r_2^2 \nu_1 \nu_3 - (\lambda-1)\nu_2 d\phi_3 \right) \ , \\
\widehat{e}^{\,2} &\equiv \widehat{e}^{\,2}_{+} = \frac{1}{ \lambda \varphi } \left( r_1^2 \varphi^2 d\phi_1 + r_2^2 ( \nu_3 - \nu_1 \nu_2 r_3^2 ) d\phi_2 - r_3^2( \nu_2 + \nu_1 \nu_3 r_2^2 ) d\phi_3 \right) \ .
\end{aligned}
\end{equation}
Following the dualisation procedure the Lorentz transformation in eq.~\eqref{eq:Lorentztrans} is given by
\begin{equation}
\Lambda = \frac{1}{\lambda} \left(\begin{array}{cc} 2-\lambda & - 2 \sqrt{\lambda - 1} \\ 2 \sqrt{\lambda -1} & 2- \lambda \end{array} \right) \ ,
\end{equation}
for which the corresponding action on spinors is simply
\begin{equation}
\Omega =\frac{1}{\sqrt{\lambda}} \mathbb{I} - \frac{\sqrt{\lambda -1 }}{\sqrt{\lambda} }\Gamma^{12} \ .
\end{equation}
Then acting on the poly-form we ascertain the T-dual fluxes
\begin{equation}
\begin{aligned}
\widehat{F}_3&= -4 e^{-\phi_0} r_1 r_2 r_3 \, e^\alpha \wedge e^\xi \wedge \left( \nu_1 d\phi_1 + \nu_2 d\phi_2 + \nu_3 d\phi_3 \right) \ , \\
\widehat{F}_5 &= (1 +\star) \frac{4 e^{-\phi_0}}{\lambda} r_1 r_2 r_3 \, e^\alpha \wedge e^\xi \wedge d\phi_1 \wedge d\phi_2 \wedge d\phi_3 \ ,
\end{aligned}
\end{equation}
in complete agreement with the results of \cite{Frolov:2005dj}.

To close this section let us make a small observation. For the $\beta$-deformation $\nu_1 = \nu_2 = \nu_3 \equiv \gamma$ there a special simplification that happens when $\gamma = \frac{1}{n}$, $n\in \mathbb{Z}$. In this case the deformed gauge theory is equivalent to that of D3 branes on the discrete torsion orbifold $\mathbb{C}^3/\Gamma$ with $\Gamma = \mathbb{Z}_n \times \mathbb{Z}_n$. These cases are also special in the dualisation procedure above. Notice that the Lagrange multiplier $v$ corresponding to the central extension is inversely proportional to $\gamma$ and hence the orbifold points correspond to cases where $v$ is integer quantised. Moreover, recalling that non-abelian T-duality with respect to a centrally-extended $U(1)^2$ is equivalent to first adding a total derivative $B$-field, i.e. making a large gauge transformation, and then T-dualising with respect to $U(1)^2$, where the required total derivative is again given by the expression in footnote~\ref{foot:bdiff}, we find that at the orbifold points ($\nu_1 = \nu_2 = \nu_3 \equiv \gamma = \frac1n$) the integral of this total derivative
 \begin{equation}
\frac1{4\pi^2} \int B_2 = \frac{n}{12\pi^2} \int (d\phi_2 \wedge d\phi_3 + d \phi_1 \wedge d \phi_2 + d\phi_3 \wedge d\phi_1) = n \ ,
\end{equation}
is also integer quantised.

\subsection{Application 3: Dipole Deformations}\label{ssec:app3}

Dipole theories \cite{Bergman:2000cw,Bergman:2001rw} are a class of non-local field theories obtained from regular (or even non-commutative) field theories by associating to each non-gauge field $\Phi_a$ a vector $L^\mu_a$ and replacing the product of fields with a non-commutative product
\begin{equation}
(\Phi_1 \tilde\star \Phi_2 )(x) \equiv \Phi_1(x- \frac{1}{2} L_2) \Phi_2 (x+ \frac{1}{2} L_1) \ .
\end{equation}
Whilst intrinsically non-local, these theories can be mapped to local field theories with a tower of higher-order corrections. For small $L$ the leading correction is the coupling to a dimension 5 operator, which for $\mathcal{N} = 4$ SYM was identified in \cite{Bergman:2000cw} as
\begin{equation}
\Delta {\cal L} = L^\mu \cdot {\cal O}_\mu \ , \quad {\cal O}_\mu^{IJ} = \frac{i}{g^2_{YM}} \textrm{tr}\left(F_\mu{}^\nu \Phi^{[I} D_{\mu} \Phi^{J]}+ (D_\mu \Phi^K)\Phi^{[K}\Phi^I\Phi^{J]} \right) \ .
\end{equation}
In \cite{Bergman:2001rw} the supergravity dual to this dipole deformation was constructed. When aligned in the $x^3$ direction the dipole vector $L$ specifies a constant element in $ \mathfrak{su}(4)$ which defines in the $\bf{4}$ a $4\times 4$ traceless hermitian matrix $U$ and in the $\bf{6}$ a $6\times 6$ real antisymmetric matrix $M$. In terms of these matrices the supergravity metric is given by \cite{Bergman:2001rw}
\begin{equation}
ds^2 = \frac{R^2}{z^2} \left( -dt^2 + dx_1^2 + dx_2^2 + f_{1}^{-1}z^2 d x_3^2 \right) + R^2 \left( d\textrm{n}^Td\textrm{n} + \lambda^2 f_{1}^{-1} (\textrm{n}^T M d \textrm{n})^2 \right) \ ,
\end{equation}
where $\textrm{n}$ is a unit vector in $\mathbb{R}^6$, $\lambda = R^4 (\alpha^\prime)^{-2}= 4 \pi g^2_{YM}N$ and
\begin{equation}
f_1 = \frac{z^2}{R^2}+ \lambda^2 \textrm{n}^T M^T M \textrm{n} \ .
\end{equation}
The deformation acts in both $S^5$ and $AdS^5$. The eigenvalues of a $6 \times 6$ real antisymmetric matrix are three imaginary numbers and their complex conjugates. If we take three of the independent eigenvalues of $M$ to be equal, $M^T M$ is a positive constant, $l^2/\lambda^2$, times the identity matrix, and hence
\begin{equation}
f_1 = z^2 + l^2 \ ,
\end{equation}
where we have set $R = 1$. Though this case preserves no supersymmetry, it does yield a simple metric on the five-sphere; viewed as a $U(1)$ fibration over $\mathbb{C}\mathbf{P}^2$ (given in appendix \ref{app:algconv} in eq.~\eqref{eq:CPform}) the deformation acts to change the radius of this fibration such that it depends on the function $f_1$ \cite{Bergman:2001rw}, which now only depends on the $AdS$ radial coordinate.

To arrive at this dipole deformation via centrally-extended non-abelian T-duality we gauge the central extension of the $U(1)^{2}$ subgroup generated by $\{ \mathfrak{P}_{3} , (S_{12}+ S_{34}+ S_{56}) \}$. We gauge fix the coset representative
\begin{equation}
\hat{g} = g_{AdS_{5}}\oplus g_{S^{5}} \ ,\quad x_{3}\rightarrow 0 \ , \quad \phi \rightarrow 0 \ ,
\end{equation}
where $g_{AdS_{5}}$ is the parametrisation relevant for the Poincar\'e patch \eqref{eq:gAdS} and $g_{S^{5}}$ is given in eq.~\eqref{eq:CPparam}. The Lagrange multipliers are then parametrised as
\begin{equation}
v_{1} = \frac{\phi}{l} \ , \quad v_{2}= \frac{x_{3}}{l} \ , \quad v_{3} = \frac{1}{l} \ .
\end{equation}
Following the general formulae one arrives at the T-dual frame fields
\begin{equation}
\widehat{e}^{\,1}_{\pm} = \frac{z}{z^{2} + l^{2}} \big( dx_{3} \pm l \Psi \big) \ , \quad \widehat{e}^{\,2}_{\pm} = \frac{z}{z^{2} + l^{2}} \big( -z \Psi \pm \frac{l}{z} dx_{3} \big) \ ,
\end{equation}
in which $\Psi$ is the global one-form corresponding to the $U(1)$ fibration defined in eq.~\eqref{eq:CPform}. It is a simple matter to extract the Lorentz rotation in the spinor representation
\begin{equation}
\Omega = \frac{1}{\sqrt{z^{2}+ l^{2}} }\left(z \mathbb{I} - l \Gamma^{12} \right) \ .
\end{equation}

Here $\Gamma_{12}$ refers to the directions in tangent space given by frames $\widehat{e}^{\,1}$ and $\widehat{e}^{\,2}$. This is a product of two gamma matrices, one with legs in $S^{5}$ and the other in $AdS_{5}$. Therefore, the action of $\Omega$ only produces a five-form in the dualised target space. In fact since, for example, $z\widehat{e}^{\,2}_+ - l \widehat{e}^{\,1}_+ = - z \Psi$ one finds that $F_5$ is only altered by an overall constant scaling that could be re-absorbed into a shift of the dilaton. The final result is the target space geometry
\begin{equation}
\begin{aligned}
\widehat{ds}^{2} &= \frac{1}{z^{2}} \left( -dx_{0}^{2}+ dx_{1}^{2}+dx_{2}^{2 } + dz^{2}\right) + ds^{2}_{\mathbb{C}\mathbf{P}^2} + \frac{1}{z^{2}+l^{2}} dx_{3}^{2 } + \frac{z^{2}}{z^{2}+l^{2}}\Psi^{2} \ , \\
\widehat{B} &= \frac{l}{z^{2}+ l^{2}}\Psi \wedge dx_{3} + \frac{1}{l} dx_{3} \wedge d\phi \ , \quad
e^{ 2(\widehat{\Phi}-\phi_{0})} = \frac{z^{2}l^{2}}{z^{2}+l^{2}} \ , \quad \widehat{F}_5 = \frac{1}{l} F_5 \ .
\end{aligned}
\end{equation}
Modulo a gauge transformation in $\widehat{B}$ this agrees with the geometry of \cite{Bergman:2001rw}.

\section{Concluding Comment}\label{sec:conclusions}

In this article we have demonstrated that the holographic dual of many known deformations of gauge theories can be understood in terms of non-abelian T-duality, extending the construction in the NS sector of \cite{Hoare:2016wsk,Borsato:2016pas} to the RR sector. In section \ref{sec:examples} we tested the construction on a number of examples: a non-commutative deformation, the $\gamma$-deformation, a dipole deformation and, in appendix \ref{app:furtherexamples}, a unimodular non-abelian deformation and a jordanian deformation.

There are a number of interesting open directions. Our construction involved only bosonic generators of the $\mathfrak{psu}(2,2|4)$ algebra of the superstring; it would be interesting to extend this to more general $r$-matrices, including those containing fermionic generators. Furthermore, to formalise the relation between the Yang-Baxter deformations and non-abelian dualities it would be useful to understand how the spinor rotation defining the deformed RR fluxes in the former \cite{Borsato:2016ose} is related to that in the latter, which was the subject of the present article. Additionally, one would like to understand whether solutions of the modified cYBE (i.e. $\eta$-deformations and their Poisson-Lie dual $\lambda$-deformations) can be understood in this framework. Finally, and perhaps optimistically, one might hope that generalised notions of T-duality can be employed to find gravitational duals of other non-integrable marginal deformations of gauge theories.

\section{Acknowledgements}

It is a pleasure to thank Saskia Demulder, Carlos N\'u\~nez, Arkady Tseytlin, Linus Wulff and Konstantinos Zoubos for discussions concerning aspects of this work.
The work of BH is partially supported by grant no. 615203 from the European Research Council under the FP7.
The work of DT is supported in part by the Belgian Federal Science Policy Office through the Interuniversity Attraction Pole P7/37, and in part by the ``FWO-Vlaanderen'' through the project G020714N and a postdoctoral fellowship, and by the Vrije Universiteit Brussel through the Strategic Research Program ``High-Energy Physics''.

\begin{appendix}

\section{(Modified) Supergravity Conventions}\label{app:sugra}

In this appendix we summarise our conventions for the (modified) Type IIB supergravity equations. Similar equations exist for Type IIA.
Let us define the following beta-functions
\begin{equation}\label{eq:betafunctions}
\begin{aligned}
\beta_{mn}^{G} &= R_{mn} + 2\nabla_{m n }\Phi -\frac{1}{4}H_{mpq}H_{n}{}^{pq}
\\ & - e^{2\Phi}\Big( \frac{1}{2}({F_1}^2)_{mn}+\frac{1}{4}({F_3}^2)_{mn} +\frac{1}{96}({F_5}^2)_{mn} - \frac{1}{4}g_{mn} \big( F_1^2 +\frac{1}{6}F_{3}^2 \big)\Big) \ , \\
\beta_{mn}^B &= d[e^{-2\Phi} \star H] - F_1\ww \star F_3 - F_3 \ww F_5 \ , \\
\beta^\Phi &= R+ 4 \nabla^2 \Phi - 4 (\partial \Phi)^2 - \frac{1}{12} H^2 \ .
\end{aligned}
\end{equation}
For a globally scale invariant $\sigma$-model the beta-functions for the metric and $B$-field vanish up to diffeomorphisms and gauge transformations. There should then be similar second-order equations for the RR fluxes.

The Type IIB supergravity equations, i.e. the critical string equations, are given by
\begin{equation}\label{eq:sugraeq}
\begin{aligned}
\beta^G_{mn} & = 0 \ , \quad \beta^B_{mn} = 0 \ , \quad \beta^\Phi = 0 \ ,
\\
d\cF_1 & = d\Phi \ww \cF_1 \ , \quad
d\star\cF_1 + H\ww \star \cF_3 = d\Phi \ww\star \cF_1 \ , \\
d\cF_3 - H\ww \cF_1 &= d\Phi \ww \cF_3 \ , \quad
d\star \cF_3 + H\ww \star \cF_5 = d\Phi \ww \star \cF_3 \ ,\\
d\cF_5 - H\ww \cF_3&= d\Phi \ww \cF_5 \ , \quad \cF_5 = \star \cF_5 \ , \\
\end{aligned}
\end{equation}
where we have defined $\cF = e^\Phi F$.

There exists a modification to the supergravity equations that still imply the global scale invariance conditions, but now depend on an additional Killing vector of the background $I$. These modified supergravity equations can be understood as follows. We start from a solution of the Type II supergravity equations for which the metric, $B$-field and weighted RR fluxes $\cF$ have a $U(1)$ isometry corresponding to shifts in the coordinate $y$, but where the dilaton breaks this isometry via a piece linear in $y$, i.e. $\Phi = cy + \ldots$. The supergravity equations only depend on $d\Phi$ and hence we can ask what happens if we formally T-dualise in $y$. The dual background then solves the modified equations with the Killing vector corresponding to shifts in the dual coordinate to $y$ \cite{Arutyunov:2015mqj}. Alternatively they follow from the requirement that the Type II Green-Schwarz string action is $\kappa$-symmetric \cite{Wulff:2016tju}. Recently they have also been formulated in an $O(d,d)$ invariant manner, as a modification of Type II double field theory \cite{Sakatani:2016fvh}. The modified Type IIB supergravity equations are
\begin{equation}\label{eq:modsugra1}
\begin{aligned}
\beta_{mn}^{G} &= - \nabla_m W_n - \nabla_n W_m \ , \quad
e^{2\Phi} \beta_{mn}^{B} = 2 \star d W +2 W \wedge \star H \ ,
\\
\beta^{\Phi} &= 4 \star d \star W -4 \star( (W+ 2 d\Phi) \ww \star W) \ ,
\\
d\cF_1 & = Z\ww \cF_1 + \star (I \ww \star \cF_3) \ , \quad
d\star\cF_1 + H\ww \star \cF_3 = Z\ww\star \cF_1 \ , \quad \star (I \ww \star \mathcal{F}_1) = 0 \ ,
\\
d\cF_3 - H\ww \cF_1 &= Z\ww \cF_3 + \star (I \ww \star \cF_5) \ , \quad
d\star \cF_3 + H\ww \star \cF_5 = Z\ww \star \cF_3- \star( I \wedge \cF_1 ) \ ,
\\
d\cF_5 - H\ww \cF_3&= Z\ww \cF_5- \star (I \ww \cF_3) \ , \quad \cF_5 = \star \cF_5 \ ,
\end{aligned}
\end{equation}
where $I$ is a one-form corresponding to a certain Killing vector of the background, i.e.
\begin{equation}\label{eq:modsugra2}
\mathcal{L}_I G= \mathcal{L}_I B = \mathcal{L}_I \Phi = \mathcal{L}_I \cF_{1,3,5} = 0 \ ,
\end{equation}
and the one-forms $Z$, $X$ and $W$ are constructed from $I$ and $\Phi$
\begin{equation}\label{eq:modsugra3}
Z = d\Phi - \iota_I B \ , \quad X= I + Z \ , \quad W = X - d\Phi = I - \iota_I B \ .
\end{equation}

It is important to note that for the modified system of equations to be invariant under the gauge freedom $B \to B + d \Lambda$ (where for simplicity we assume that $\mathcal{L}_I \Lambda = 0$) the ``dilaton'' field $\Phi$ must now transform as $\Phi \to \Phi - \iota_I \Lambda$, and hence is not unique. This can be understood by starting from a Weyl-invariant background with a dilaton linear in an isometric direction $y$, $\Phi = c y + \ldots$. If we shift $y$ by an arbitrary function of the transverse coordinates this ansatz is preserved, however the explicit form of the dilaton is changed. After ``T-dualising'' in $y$ this coordinate redefinition then maps to a gauge transformation under which the dual ``dilaton'' field now transforms.

\section{Conventions for Algebras}\label{app:algconv}
In this appendix we outline our conventions for the algebras $\mathfrak{so}(4,2)$ and $\mathfrak{so}(6)$ for which we largely adopt those of \cite{Arutyunov:2009ga}.
For $SO(4,2)$ we start by defining the $\gamma$ matrices
\begin{equation}
\gamma_0 = i \sigma_3 \otimes \sigma_0 \ , \quad \gamma_1= \sigma_2 \otimes \sigma_2 \ , \quad \gamma_2= -\sigma_2 \otimes \sigma_1\ , \quad \gamma_3= -\sigma_1 \otimes \sigma_0 \ , \quad \gamma_4= \sigma_2 \otimes \sigma_3 \ ,
\end{equation}
in terms of which the generators of $SO(4,2)$ are given by
\begin{equation}
T_{ij} = \frac{1}{4} [\gamma_{i}, \gamma_j] \ , \quad T_{i5} = \frac{1}{2} \gamma_i \ , \quad i,j = 0, \ldots, 4 \ .
\end{equation}
The $SO(4,1)$ subgroup is generated by $T_{ij}$ for $i,j = 0, \ldots, 4$. The projector onto the orthogonal complement is given by
\begin{equation}
P(X)= - \Tr(X T_{0,5}) T_{0,5} + \sum_{i=1}^4 \Tr(X T_{i,5}) T_{i,5} \ .
\end{equation}

A useful adapted basis when considering Poincar\'e patch is
\begin{equation}
\mathfrak{D}= T_{45} \ , \quad \mathfrak{P}_\mu = T_{\mu 5} - T_{\mu 4} \ , \quad \mathfrak{K}_\mu = T_{\mu 5} + T_{\mu 4} \ , \quad \mathfrak{M}_{\mu \nu} = T_{\mu \nu} \ , \quad \mu = 0,\ldots, 3 \ .
\end{equation}
We also use $\mathfrak{M}_{+i} = \mathfrak{M}_{0i} +\mathfrak{M}_{1i} $ for $i=2,3$.
The bosonic $AdS_5$ $\sigma$-model is given by
\begin{equation}
{\cal L } = \Tr(J_+ P(J_-) ) \ ,
\end{equation}
for $J= g^{-1}dg$ and when the gauged fixed group element is parametrised as
\begin{equation}
\label{eq:gAdS}
g= \exp\left[ \eta^{\mu \nu} x_\mu \mathfrak{P}_\nu \right] z^{\mathfrak{D} } \ ,
\end{equation}
the target space metric is given on the Poincar\'e patch by
\begin{equation}
ds^2=\frac{1}{z^2} \left( dz^2 + \eta^{\mu \nu} dx_\mu dx_\nu \right) \ .
\end{equation}
As usual the coordinate $u$ used in section \ref{ssec:app1} is related to $z$ by $u = z^{-1}$. In the examples that we consider we dualise with respect to a subalgebra $\mathfrak{h}\subset \mathfrak{so}(4,2)$ which does not necessarily need to be a subgroup of the $\mathfrak{so}(4,1)$ subalgebra specified above.

For $\mathfrak{so}(6)\cong \mathfrak{su}(4)$ we supplement $\gamma_i$ $i=1,\ldots, 4$, defined above with $\gamma_5 = -i \gamma_0$ and construct the (anti-hermitian) generators
\begin{equation}
S_{ij} = \frac{1}{4} [\gamma_{i}, \gamma_j] \ , \quad S_{i6} = \frac{i}{2} \gamma_i \ , \quad i,j = 1, \ldots, 5 \ .
\end{equation}
The Cartan subalgebra is generated by
\begin{equation}
h_1= i \diag(1,1,-1,-1) \ , \quad h_2= i \diag(1,-1,1,-1) \ , \quad h_3= i \diag(1,-1,-1,1)
\end{equation}
We take the $\mathfrak{so}(5)$ subalgebra to be generated by $S_{ij}$ for $i=1,\ldots,5$, such that the projector onto the orthogonal complement of this subgroup is
\begin{equation}
P(X) = \sum_{i=1}^5\Tr( X\cdot S_{i6}) S_{i6} \ ,
\end{equation}
where here $\Tr$ stands for the negative of the matrix trace. A coset representative for $SO(6)/SO(5)$ can be chosen as
\begin{equation}
g = \exp[\frac{1}{2} \phi^m h_m ] \exp[-\frac{\xi}{2} \gamma^{13}] \exp[\frac{i\alpha}{2} \gamma^1] \ ,
\end{equation}
leading to the $\sigma$-model parametrisation of $S^5$ employed in section \ref{ssec:app2}.

An alternative parametrisation is given by
\begin{equation}
\label{eq:CPparam}
g= \exp[\frac{i\phi}{2} \gamma_5] \cdot \big[s \, \mathbb{I} + \frac{it}{2} \big( e^{i\phi} \alpha (\gamma_1 - i \gamma_2) + e^{-i\phi} \bar\alpha (\gamma_1 + i \gamma_2) + e^{i\phi} \beta (\gamma_3- i \gamma_4) + e^{-i\phi}\bar\beta (\gamma_3+ i \gamma_4) \big) \big] \ ,
\end{equation}
where
\begin{equation}
r = 1+ |\alpha|^2 + |\beta|^2 \ , \quad s^2 = \frac{1}{2\sqrt{r} }(1+ \sqrt{r}) \ , \quad t^2 = \frac{1}{2\sqrt{r} (1+ \sqrt{r})} \ .
\end{equation}
These coordinates give a metric on $S^5$ that makes manifest the structure of $S^5$ as a $U(1)$ fibration over $\mathbb{C}\mathbf{P}^2$
\begin{equation}
\label{eq:CPform}
\begin{aligned}
&ds^2_{S^5} = ds^{2}_{\mathbb{C}\mathbf{P}^2} + \Psi^{2}\ , \quad ds^{2}_{\mathbb{C}\mathbf{P}^2} = \frac{1}{r}( |d\alpha|^2 + |d\beta|^2 ) - \frac{1}{r^2}| \omega |^2 \ , \\
& \Psi= d\phi + \frac{1}{r} \Im( \omega) \ , \quad \omega = \bar\alpha d \alpha + \bar\beta d \beta \ .
\end{aligned}
\end{equation}
The global one-form $\Psi = \sum_{i=1\dots 3} x_i dy_i - y_i d x_i$ where $z_i = x_i + i y_i$ are coordinates on $\mathbb{C}^3$ given by $z_1 = \frac{1}{\sqrt{r}} e^{i \phi}$, $z_2 = \frac{\alpha}{\sqrt{r}} e^{i \phi}$, $z_3 = \frac{\beta}{\sqrt{r}} e^{i \phi}$. One can think of $\Psi$ as a contact form whose corresponding Reeb vector has orbits which are the $S^1$ fibres. For computational purposes we note that frame fields for $\mathbb{C}\mathbf{P}^2$ can be found in e.g. \cite{Eguchi:1980jx}.

When dealing with the dipole deformation in section \ref{ssec:app3} we will need the full ten-dimensional space-time. This is readily achieved by taking a block diagonal decomposition, i.e. $g = g_{AdS_{5}} \oplus g_{S^5}$, with the generators of $\mathfrak{su}(2,2) \oplus \mathfrak{su}(4)$ given by $8 \times 8$ matrices, with the $\mathfrak{su}(2,2)$ and $\mathfrak{su}(4)$ generators in the upper left and lower right $4 \times 4$ blocks respectively. Traces are then replaced with ``supertrace'' (the bosonic restriction of the supertrace on $\mathfrak{psu}(2,2|4)$) given by the matrix trace of the upper $\mathfrak{su}(2,2)$ block minus the matrix trace of the lower $\mathfrak{su}(4)$ block.

\section{Further Examples of Deformations in \texorpdfstring{$AdS_5$}{AdS5}}\label{app:furtherexamples}

In section \ref{sec:examples} we considered non-abelian T-dualities with respect to a centrally-extended two-dimensional abelian algebra, demonstrating that this is equivalent to a TsT transformation of the full supergravity background. There are additional classes of deformations that can be constructed as non-abelian T-duals. These come from considering particular non-semisimple subalgebras of $\mathfrak{su}(2,2) \oplus \mathfrak{su}(4)$, whose existence relies on the non-compactness of $\mathfrak{su}(2,2)$. There are a number of such algebras that are non-abelian and admit central extensions \cite{Borsato:2016ose}, such that when we T-dualise the metric with respect to this centrally-extended subalgebra we find a deformation of the original metric \cite{Hoare:2016wsk,Borsato:2016pas} that coincides with a certain Yang-Baxter deformation. To illustrate this richer story we present a summary of two examples showing how the techniques described in this paper also apply, i.e. the R-R fluxes following from non-abelian T-duality agree with those of the Yang-Baxter $\sigma$-model.

An $r$-matrix
\begin{equation}
r = r^{ab} T_a \wedge T_b \ ,
\end{equation}
is said to be non-abelian if $[T_a , T_b]\neq 0$ for at least some of the generators. An $r$-matrix is said to be unimodular if
\begin{equation}
r^{ab} [T_a, T_b] = 0 \ .
\end{equation}
For a solution of the classical Yang-Baxter equation the unimodularity of the $r$-matrix is equivalent to the unimodularity ($f_{ab}{}^b = 0$) of the corresponding subalgebra. In \cite{Borsato:2016ose} it was shown that the background defined by a Yang-Baxter $\sigma$-model based on a non-unimodular non-abelian $r$-matrix is not a supergravity background, but rather solves the modified supergravity described above. The first example we discuss corresponds to a non-abelian but unimodular $r$-matrix, while the second is a non-unimodular $r$-matrix.

\subsection{Unimodular \texorpdfstring{$r$}{r}-matrix}\label{sapp:340}

The first example corresponds to an $r$-matrix considered in \cite{Borsato:2016ose}
\begin{equation}
r = \eta~ \mathfrak{M}_{23}\wedge \mathfrak{P}_1 + \zeta~ \mathfrak{P}_2\wedge \mathfrak{P}_3 \ .
\end{equation}
This is non-abelian e.g. $[\mathfrak{M}_{23}, \mathfrak{P}_2]= - \mathfrak{P}_3$, but since $[\mathfrak{M}_{23}, \mathfrak{P}_1]= [\mathfrak{P}_{2}, \mathfrak{P}_3]=0$ it is unimodular. In \cite{Borsato:2016ose} it was shown that the corresponding deformation is nevertheless equivalent to two non-commuting TsT transformations, with a non-linear coordinate redefinition in between. On the other hand it was discussed from the perspective of non-abelian T-duality in \cite{Hoare:2016wsk} where the relevant subalgebra was $\mathfrak{h}= \{ \mathfrak{M}_{23}, \mathfrak{P}_1 , \mathfrak{P}_2, \mathfrak{P}_3 \}$. The gauge freedom can be used to fix the coset representative in eq.~\eqref{eq:gAdS} to $\hat{g} = e^{-x_{0 }\mathfrak{P}_{0}} z^{\mathfrak{D}}$, but there remains one residual gauge symmetry which is used to fix a Lagrange multiplier to zero. The Lagrange multipliers are parametrised by
\begin{equation}
v_1= -\frac{x_1}{\eta} + \frac{r^2}{2 \zeta} \ , \quad v_2= \frac{\theta}{\eta}\ , \quad v_3=\frac{r}{\zeta}\ ,\quad v_4= 0\ , \quad v_5= \frac{1}{\eta}\ , \quad v_6= \frac{1}{\zeta} \ ,
\end{equation}
where $v_5$ and $v_6$ correspond to the two central generators and $r$ and $\theta$ are polar coordinates on the $x_2 ,x_3$ plane. Applying the non-abelian T-duality technology one finds the dual geometry is
\begin{equation}
\begin{aligned}
\widehat{ds}^2 &=\frac{1}{z^2} \left( dz^2 - dx_0^2\right) + \widehat{e}_\pm \cdot \widehat{e}_\pm + ds^2_{S^5} \ ,
\\
\widehat{e}^{\,1}_+ &= \frac{dx_1 \left(\zeta ^2+z^4\right)+\eta r \left(-\zeta dr +r z^2 d\theta \right)}{z f } \ ,
\\ \widehat{e}^{\,2}_+& = \frac{z \left(\zeta dr+\eta r dx_1-r z^2 d \theta
\right)}{f} \ , \\
\widehat{e}^{\,3}_+&= \frac{-dr \left(\eta ^2 r^2+z^4\right)+\zeta \eta r
dx_1-\zeta r z^2 d\theta }{z f}\ ,
\end{aligned}
\end{equation}
where $f= \zeta ^2+\eta ^2 r^2+z^4$, while the remaining NS fields are
\begin{equation}
\widehat{B} = \frac{-\zeta \eta r dr \wedge d\theta + \left(\zeta ^2+z^4\right) dx_1\wedge d \theta }{\eta f } \ , \quad
e^{-2(\widehat{\Phi} - \phi_0)} = \frac{f}{\zeta ^2 \eta ^2 z^4} \ .
\end{equation}
The Lorentz rotation $\Lambda e_- = e_+$ is given by
\begin{equation}
\Lambda = \frac{1}{f} \left(
\begin{array}{ccc}
z^4+\zeta ^2-r^2 \eta ^2 & -2 r z^2 \eta
& 2 r \zeta \eta \\
2 r z^2 \eta & z^4-\zeta ^2-r^2 \eta
^2 & - 2 z^2 \zeta \\
2 r \zeta \eta & 2 z^2 \zeta & z^4-\zeta ^2+r^2 \eta ^2 \\
\end{array}
\right) \ ,
\end{equation}
with the corresponding spinor representation
\begin{equation}
\Omega = \frac{1}{\sqrt{f} } \left( z^2 \mathbb{I} - r \eta \Gamma^{12} - \zeta \Gamma^{23} \right) \ .
\end{equation}
This completes the IIB supergravity solution with the three-form and five-form flux
\begin{equation}
\begin{aligned}
F_3 &=\frac{4 e^{-\phi_0} }{z^5 \zeta \eta }\left( \zeta dx_0\wedge dx_1 \wedge dz - r \eta dx_0 \wedge dr \wedge dz \right) \ , \\
F_5&= (1+\star) \frac{-4 e^{-\phi_0} r }{ z \zeta \eta f} dx_0\wedge dx_1 \wedge dr \wedge dz\wedge d\theta \ ,
\end{aligned}
\end{equation}
in agreement with the expressions following from the Yang-Baxter $\sigma$-model \cite{Borsato:2016ose}.

\subsection{Non-unimodular \texorpdfstring{$r$}{r}-matrix}\label{sapp:366a}

The final example we consider is an $r$-matrix that can be found by infinitely boosting the Drinfel'd-Jimbo solution to the modified classical Yang-Baxter equation for $\mathfrak{su}(2,2)$ \cite{Hoare:2016hwh}
\begin{equation}\label{eq:rmatnm}
r= \eta \left(\mathfrak{D} \wedge \mathfrak{P}_0 + \mathfrak{M}_{01}\wedge \mathfrak{P}_1 + \mathfrak{M}_{+2}\wedge \mathfrak{P}_2 + \mathfrak{M}_{+3}\wedge \mathfrak{P}_3 \right) \ .
\end{equation}
This $r$-matrix of jordanian type and the corresponding deformations of the $AdS_5 \times S^5$ superstring were first studied in \cite{Kawaguchi:2014qwa,Kawaguchi:2014fca}. Furthermore, the $r$-matrix is non-unimodular and the corresponding dualisation of $AdS_5$ with respect to the non-abelian subalgebra
\begin{equation}
\mathfrak{h}= \{\mathfrak{D} , \mathfrak{P}_0 , \mathfrak{M}_{01}, \mathfrak{P}_1, \mathfrak{M}_{+2} , \mathfrak{P}_2 , \mathfrak{M}_{+3} , \mathfrak{P}_3\} \ ,
\end{equation}
is afflicted with a mixed gravity/gauge anomaly (i.e. $n_a= f_{ab}{}^b \neq 0$) \cite{Elitzur:1994ri}. The algebra $\mathfrak{h}$ admits a single central extension with the commutator of each pair of generators in \eqref{eq:rmatnm} being extended by the same generator. Since all directions are dualised the coset representative is fully fixed to $\hat{g}=1$ leaving three further gauge fixings to be made on the dynamical Lagrange multipliers. We parametrise these as
\begin{equation}
v_1= \frac{x_0}{\eta} \ , \quad v_2= \frac{-1+z}{\eta} \ , \quad v_3 = \frac{x_1}{\eta} \ , \quad v_5 + i v_7 = \frac{r e^{i\theta}}{\eta} \ , \quad v_4=v_6=v_8=0 \ , \quad v_9 = \frac{1}{\eta} \ ,
\end{equation}
where $v_9$ corresponds to the central direction. The dual metric is given by
\begin{equation}
\begin{aligned}
&\widehat{ds}^2 =e_\pm^i \eta_{ij} e_\pm^i + \widehat{ds}^2_{S^5} \ , \quad \eta_{ij} = \textrm{diag} (1,-1,1, 1,1) \ , \\
&\widehat{e}^{\,1}_+= \frac{1}{p}(-\eta dx_0 + z dz) \ , \quad \widehat{e}^{\,2}_+= \frac{1}{p}(-z dx_0 + \eta dz) \ , \quad \widehat{e}^{\,3}_+= -\frac{z}{q}(z^2 dx_1 + r \eta dr) \ ,\\
& \widehat{e}^{\,4}_+ + i \widehat{e}^{\,5}_+ = \frac{e^{i\theta}}{q}\left( r z \eta dx_1 -z^3 dr - \frac{i q r}{z} d\theta \right) \ ,
\end{aligned}
\end{equation}
where $p= z^2-\eta^2$ and $q= z^4+ r^2 \eta^2$. The remaining NS fields are
\begin{equation}
\widehat{B} = \frac{z}{p \eta}dz\wedge dx_0 + \frac{r\eta}{q} dr \wedge dx_1 \ , \quad e^{-2(\widehat{\Phi}-\phi_0)} = \frac{p q z^2}{\eta^8} \ .
\end{equation}

The $SO(1,4)$ Lorentz rotation has a block diagonal decomposition $\Lambda = \Lambda_1 \oplus \Lambda_2$ with
\begin{equation}
\Lambda_1 = \frac{1}{p}\left(\begin{array}{cc} z^2+\eta^2 & 2 z \eta \\ 2 z \eta & z^2 +\eta^2 \end{array} \right) \ , \ \ \
\Lambda_2= \frac{1}{q}\left(\begin{array}{ccc} z^4 -r^2 \eta^2 & 2 r z^2 \eta {\cal C}_\theta & 2 r z^2 \eta {\cal S}_\theta \\
-2 r z^2 \eta {\cal C}_\theta & z^4 - r^2 \eta^2 {\cal C}_{2\theta} & -r^2 \eta^2 {\cal S}_{2\theta} \\
-2 r z^2 \eta {\cal S}_\theta &- r^2 \eta^2 {\cal S}_{2\theta} & z^4 + r^2 \eta^2 {\cal C}_{2\theta}
\end{array}\right) \ .
\end{equation}
The corresponding spinor rotation $\Omega = \Omega_1\cdot\Omega_2$ is given by (recalling the signature is such that $(\Gamma^{2})^{2}= - \mathbb{I} $ whilst the remaining $(\Gamma^{i})^{2}= \mathbb{I}$)
\begin{equation}
\begin{aligned}
\Omega_1 = \frac{1}{\sqrt{p}} \left( z \mathbb{I} + \eta \Gamma^{12} \right) \ , \quad
\Omega_2 = \frac{1}{\sqrt{q}} \left(z^2 \mathbb{I} + r\eta \cos\theta \Gamma^{34} + r\eta \sin\theta \Gamma^{35} \right) \ .
\end{aligned}
\end{equation}
This gives the fluxes
\begin{equation}
\begin{aligned}
F_1&= \frac{ 4e^{-\phi_0}}{\eta^2} r^2 d\theta \ ,
\quad
F_3 = \frac{4e^{-\phi_0} r z^4}{\eta^3 q } dx_1\wedge dr\wedge d\theta - \frac{ 4e^{-\phi_0} r^2 z}{\eta^3 p} dx_0\wedge dz \wedge d\theta \ , \\
F_5&= (1+\star) \frac{ - 4e^{-\phi_0} r z^5}{\eta^4 q p} dx_0\wedge dx_1\wedge dr \wedge dz \wedge d\theta \ .
\end{aligned}
\end{equation}
These fluxes do not solve their Bianchi identities, nor their equations of motions. Instead they solve the generalised supergravity equations above with the modification determined by the one-form $W$ given by the push forward of the worldsheet gauge field $A_+$ as in eq.~\eqref{eq:Ampush}, which in turn, via eq.~\eqref{eq:modsugra3}, yields
\begin{equation}
I = 4\frac{\eta}{p} dx_0 - 2 \frac{z^2 \eta }{q} dx_1 \ .
\end{equation}

The expressions for the metric, $e^{\widehat{\Phi}} F$ and $I$ agree with those of the background presented in \cite{Orlando:2016qqu}. Recalling that the ``dilaton'' field now transforms under the gauge freedom $B \to B + d\Lambda$, we also find that the ``dilaton'' and $B$-field match up to a gauge transformation.
\end{appendix}




\begin{thebibliography}{99}


\bibitem{Lunin:2005jy} O.~Lunin and J.~M.~Maldacena,
``Deforming field theories with $U(1) \times U(1)$ global symmetry and their gravity duals,''
JHEP {\bf 0505} (2005) 033
[\arxivlink{hep-th/0502086}].

\bibitem{Berkovits:2008ic} N.~Berkovits and J.~Maldacena,
``Fermionic T-Duality, Dual Superconformal Symmetry, and the Amplitude/Wilson Loop Connection,''
JHEP {\bf 0809} (2008) 062
[\arxivlink{0807.3196}].

\bibitem{Beisert:2008iq} N.~Beisert, R.~Ricci, A.~A.~Tseytlin and M.~Wolf,
``Dual Superconformal Symmetry from $AdS_5 \times S^5$ Superstring Integrability,''
Phys.\ Rev.\ D {\bf 78} (2008) 126004
[\arxivlink{0807.3228}].

\bibitem{Sfetsos:2010uq} K.~Sfetsos and D.~C.~Thompson,
``On non-abelian T-dual geometries with Ramond fluxes,''
Nucl.\ Phys.\ B {\bf 846} (2011) 21
[\arxivlink{1012.1320}].

\bibitem{Lozano:2016kum} Y.~Lozano and C.~N\'u\~nez,
``Field theory aspects of non-Abelian T-duality and $\mathcal{N} = 2$ linear quivers,''
JHEP {\bf 1605} (2016) 107
[\arxivlink{1603.04440}].

\bibitem{Hoare:2016wsk} B.~Hoare and A.~A.~Tseytlin,
``Homogeneous Yang-Baxter deformations as non-abelian duals of the $AdS_5$ sigma-model,''
J.\ Phys.\ A {\bf 49} (2016) no.49, 494001
[\arxivlink{1609.02550}].

\bibitem{Borsato:2016pas} R.~Borsato and L.~Wulff,
``Integrable deformations of T-dual $\sigma$ models,''
[\arxivlink{1609.09834}].

\bibitem{Klimcik:2002zj} C.~Klimcik,
``Yang-Baxter $\sigma$-models and dS/AdS T duality,''
JHEP {\bf 0212}, 051 (2002)
[\arxivlink{hep-th/0210095}].

\bibitem{Klimcik:2008eq} C.~Klimcik,
``On integrability of the Yang-Baxter $\sigma$-model,''
J.\ Math.\ Phys.\ {\bf 50}, 043508 (2009)
[\arxivlink{0802.3518}].

\bibitem{Klimcik:2014bta} C.~Klimcik,
``Integrability of the bi-Yang-Baxter $\sigma$-model,''
Lett.\ Math.\ Phys.\ {\bf 104}, 1095 (2014)
[\arxivlink{1402.2105}].

\bibitem{Delduc:2013fga} F.~Delduc, M.~Magro and B.~Vicedo,
``On classical $q$-deformations of integrable $\sigma$-models,''
JHEP {\bf 1311} (2013) 192
[\arxivlink{1308.3581}].

\bibitem{Delduc:2013qra} F.~Delduc, M.~Magro and B.~Vicedo,
``An integrable deformation of the $AdS_5 \times S^5$ superstring action,''
Phys.\ Rev.\ Lett.\ {\bf 112}, no. 5, 051601 (2014)
[\arxivlink{1309.5850}].

\bibitem{Metsaev:1998it} R.~R.~Metsaev and A.~A.~Tseytlin,
``Type IIB superstring action in $AdS_5 \times S^5$ background,''
Nucl.\ Phys.\ B {\bf 533} (1998) 109
[\arxivlink{hep-th/9805028}].

\bibitem{Berkovits:1999zq} N.~Berkovits, M.~Bershadsky, T.~Hauer, S.~Zhukov and B.~Zwiebach,
``Superstring theory on $AdS_2 \times S^2$ as a coset supermanifold,''
Nucl.\ Phys.\ B {\bf 567} (2000) 61
[\arxivlink{hep-th/9907200}].

\bibitem{Delduc:2014kha} F.~Delduc, M.~Magro and B.~Vicedo,
``Derivation of the action and symmetries of the $q$-deformed $AdS_5 \times S^5$ superstring,''
JHEP {\bf 1410} (2014) 132
[\arxivlink{1406.6286}].

\bibitem{Arutyunov:2013ega} G.~Arutyunov, R.~Borsato and S.~Frolov,
``S-matrix for strings on $\eta$-deformed $AdS_5 \times S^5$,''
JHEP {\bf 1404}, 002 (2014)
[\arxivlink{1312.3542}].

\bibitem{Arutyunov:2015qva} G.~Arutyunov, R.~Borsato and S.~Frolov,
``Puzzles of $\eta$-deformed $AdS_5 \times S^5$,''
JHEP {\bf 1512} (2015) 049
[\arxivlink{1507.04239}].

\bibitem{Hoare:2015gda} B.~Hoare and A.~A.~Tseytlin,
``On integrable deformations of superstring sigma models related to $AdS_n \times S^n$ supercosets,''
Nucl.\ Phys.\ B {\bf 897} (2015) 448
[\arxivlink{1504.07213}].

\bibitem{Hoare:2015wia} B.~Hoare and A.~A.~Tseytlin,
``Type IIB supergravity solution for the T-dual of the $\eta$-deformed $AdS_5 \times S^5$ superstring,''
JHEP {\bf 1510} (2015) 060
[\arxivlink{1508.01150}].

\bibitem{Arutyunov:2015mqj} G.~Arutyunov, S.~Frolov, B.~Hoare, R.~Roiban and A.~A.~Tseytlin,
``Scale invariance of the $\eta$-deformed $AdS_5 \times S^5$ superstring, T-duality and modified type II equations,''
Nucl.\ Phys.\ B {\bf 903} (2016) 262
[\arxivlink{1511.05795}].

\bibitem{Vicedo:2015pna} B.~Vicedo,
``Deformed integrable $\sigma$-models, classical R-matrices and classical exchange algebra on Drinfel'd doubles,''
J.\ Phys.\ A {\bf 48} (2015) no.35, 355203
[\arxivlink{1504.06303}].

\bibitem{Sfetsos:2015nya} K.~Sfetsos, K.~Siampos and D.~C.~Thompson,
``Generalised integrable $\lambda$- and $\eta$-deformations and their relation,''
Nucl.\ Phys.\ B {\bf 899} (2015) 489
[\arxivlink{1506.05784}].

\bibitem{Klimcik:2015gba} C.~Klimcik,
``$\eta$ and $\lambda$ deformations as ${\cal E}$-models,''
Nucl.\ Phys.\ B {\bf 900} (2015) 259
[\arxivlink{1508.05832}].

\bibitem{Klimcik:2016rov} C.~Klimcik,
``Poisson-Lie T-duals of the bi-Yang-Baxter models,''
Phys.\ Lett.\ B {\bf 760} (2016) 345
[\arxivlink{1606.03016}].

\bibitem{Delduc:2016ihq} F.~Delduc, S.~Lacroix, M.~Magro and B.~Vicedo,
``On $q$-deformed symmetries as Poisson-Lie symmetries and application to Yang-Baxter type models,''
J.\ Phys.\ A {\bf 49} (2016) no.41, 415402
[\arxivlink{1606.01712}].

\bibitem{Sfetsos:2013wia} K.~Sfetsos,
``Integrable interpolations: From exact CFTs to non-abelian T-duals,''
Nucl.\ Phys.\ B {\bf 880}, 225 (2014)
[\arxivlink{1312.4560}].

\bibitem{Hollowood:2014rla} T.~J.~Hollowood, J.~L.~Miramontes and D.~M.~Schmidtt,
``Integrable Deformations of Strings on Symmetric Spaces,''
JHEP {\bf 1411} (2014) 009
[\arxivlink{1407.2840}].

\bibitem{Hollowood:2014qma} T.~J.~Hollowood, J.~L.~Miramontes and D.~M.~Schmidtt,
``An Integrable Deformation of the $AdS_5 \times S^5$ Superstring,''
J.\ Phys.\ A {\bf 47} (2014) 49, 495402
[\arxivlink{1409.1538}].

\bibitem{Sfetsos:2014cea} K.~Sfetsos and D.~C.~Thompson,
``Spacetimes for $\lambda$-deformations,''
JHEP {\bf 1412}, 164 (2014)
[\arxivlink{1410.1886}].

\bibitem{Demulder:2015lva} S.~Demulder, K.~Sfetsos and D.~C.~Thompson,
``Integrable $\lambda$-deformations: Squashing Coset CFTs and $AdS_5\times S^5$,''
JHEP {\bf 1507} (2015) 019
[\arxivlink{1504.02781}].

\bibitem{Borsato:2016zcf} R.~Borsato, A.~A.~Tseytlin and L.~Wulff,
``Supergravity background of $\lambda$-deformed model for $AdS_2 \times S^2$ supercoset,''
Nucl.\ Phys.\ B {\bf 905} (2016) 264
[\arxivlink{1601.08192}].

\bibitem{Chervonyi:2016ajp} Y.~Chervonyi and O.~Lunin,
``Supergravity background of the $\lambda$-deformed $AdS_3 \times$ S$^3$ supercoset,''
Nucl.\ Phys.\ B {\bf 910} (2016) 685
[\arxivlink{1606.00394}].

\bibitem{Hollowood:2015dpa} T.~J.~Hollowood, J.~L.~Miramontes and D.~M.~Schmidtt,
``S-Matrices and Quantum Group Symmetry of k-Deformed Sigma Models,''
J.\ Phys.\ A {\bf 49} (2016) no.46, 465201
[\arxivlink{1506.06601}].

\bibitem{Kawaguchi:2014qwa} I.~Kawaguchi, T.~Matsumoto and K.~Yoshida,
``Jordanian deformations of the $AdS_5 \times S^5$ superstring,''
JHEP {\bf 1404} (2014) 153
[\arxivlink{1401.4855}].

\bibitem{Matsumoto:2014nra} T.~Matsumoto and K.~Yoshida,
``Lunin-Maldacena backgrounds from the classical Yang-Baxter equation - towards the gravity/CYBE correspondence,''
JHEP {\bf 1406} (2014) 135
[\arxivlink{1404.1838}].

\bibitem{Matsumoto:2015uja} T.~Matsumoto and K.~Yoshida,
``Schr\"odinger geometries arising from Yang-Baxter deformations,''
JHEP {\bf 1504} (2015) 180
[\arxivlink{1502.00740}].

\bibitem{Matsumoto:2014gwa} T.~Matsumoto and K.~Yoshida,
``Integrability of classical strings dual for noncommutative gauge theories,''
JHEP {\bf 1406} (2014) 163
[\arxivlink{1404.3657}].

\bibitem{Matsumoto:2015jja} T.~Matsumoto and K.~Yoshida,
``Yang-Baxter $\sigma$-models based on the CYBE,''
Nucl.\ Phys.\ B {\bf 893} (2015) 287
[\arxivlink{1501.03665}].

\bibitem{vanTongeren:2015soa} S.~J.~van Tongeren,
``On classical Yang-Baxter based deformations of the $AdS_5 \times S^5$ superstring,''
JHEP {\bf 1506} (2015) 048
[\arxivlink{1504.05516}].

\bibitem{Kyono:2016jqy} H.~Kyono and K.~Yoshida,
``Supercoset construction of Yang-Baxter deformed $AdS_5 \times S^5$ backgrounds,''
Prog.\ Theor.\ Exp.\ Phys.\ (2016) 083B03
[\arxivlink{1605.02519}].

\bibitem{Osten:2016dvf} D.~Osten and S.~J.~van Tongeren,
``abelian Yang-Baxter Deformations and TsT transformations,''
[\arxivlink{1608.08504}].

\bibitem{vanTongeren:2015uha} S.~J.~van Tongeren,
``Yang-Baxter deformations, AdS/CFT, and twist-noncommutative gauge theory,''
Nucl.\ Phys.\ B {\bf 904} (2016) 148
[\arxivlink{1506.01023}].

\bibitem{vanTongeren:2016eeb} S.~J.~van Tongeren,
``Almost abelian twists and AdS/CFT,''
[\arxivlink{1610.05677}].

\bibitem{Lozano:2011kb} Y.~Lozano, E.~\'O~Colg\'ain, K.~Sfetsos and D.~C.~Thompson,
``Non-abelian T-duality, Ramond Fields and Coset Geometries,''
JHEP {\bf 1106} (2011) 106
[\arxivlink{1104.5196}].

\bibitem{Itsios:2013wd} G.~Itsios, C.~N\'u\~nez, K.~Sfetsos and D.~C.~Thompson,
``Non-abelian T-duality and the AdS/CFT correspondence: new $\mathcal{N}=1$ backgrounds,''
Nucl.\ Phys.\ B {\bf 873} (2013) 1
[\arxivlink{1301.6755}].

\bibitem{Hassan:1999bv} S.~F.~Hassan,
``T duality, space-time spinors and RR fields in curved backgrounds,''
Nucl.\ Phys.\ B {\bf 568} (2000) 145
[\arxivlink{9907152}].

\bibitem{Benichou:2008it} R.~Benichou, G.~Policastro and J.~Troost,
``T-duality in Ramond-Ramond backgrounds,''
Phys.\ Lett.\ B {\bf 661} (2008) 192
[\arxivlink{0801.1785}].

\bibitem{Sfetsos:2010xa} K.~Sfetsos, K.~Siampos and D.~C.~Thompson,
``Canonical pure spinor (Fermionic) T-duality,''
Class.\ Quant.\ Grav.\ {\bf 28} (2011) 055010
[\arxivlink{1007.5142}].

\bibitem{Kelekci:2014ima} \"O.~Kelekci, Y.~Lozano, N.~T.~Macpherson and E.~\'O.~Colg\'ain,
``Supersymmetry and non-Abelian T-duality in type II supergravity,''
Class.\ Quant.\ Grav.\  {\bf 32} (2015) no.3,  035014
[\arxivlink{1409.7406}]. 

\bibitem{Alvarez:1994np} E.~Alvarez, L.~Alvarez-Gaume and Y.~Lozano,
``On non-abelian duality,''
Nucl.\ Phys.\ B {\bf 424} (1994) 155
[\arxivlink{hep-th/9403155}].

\bibitem{Elitzur:1994ri} S.~Elitzur, A.~Giveon, E.~Rabinovici, A.~Schwimmer and G.~Veneziano,
``Remarks on non-abelian duality,''
Nucl.\ Phys.\ B {\bf 435} (1995) 147
[\arxivlink{hep-th/9409011}].

\bibitem{Buscher:1987qj} T.~H.~Buscher,
``Path Integral Derivation of Quantum Duality in Nonlinear Sigma Models,''
Phys.\ Lett.\ B {\bf 201} (1988) 466.

\bibitem{Wulff:2016tju} L.~Wulff and A.~A.~Tseytlin,
``Kappa-symmetry of superstring sigma model and generalized 10d supergravity equations,''
JHEP {\bf 1606} (2016) 174
[\arxivlink{1605.04884}].

\bibitem{Sakatani:2016fvh} Y.~Sakatani, S.~Uehara and K.~Yoshida,
``Generalized gravity from modified DFT,''
[\arxivlink{1611.05856}].

\bibitem{Hashimoto:1999ut} A.~Hashimoto and N.~Itzhaki,
``Noncommutative Yang-Mills and the AdS/CFT correspondence,''
Phys.\ Lett.\ B {\bf 465} (1999) 142
[\arxivlink{hep-th/9907166}].

\bibitem{Maldacena:1999mh} J.~M.~Maldacena and J.~G.~Russo,
``Large N limit of noncommutative gauge theories,''
JHEP {\bf 9909} (1999) 025
[\arxivlink{hep-th/9908134}].

\bibitem{Leigh:1995ep} R.~G.~Leigh and M.~J.~Strassler,
``Exactly marginal operators and duality in four-dimensional $\mathcal{N} = 1$ supersymmetric gauge theory,''
Nucl.\ Phys.\ B {\bf 447} (1995) 95
[\arxivlink{hep-th/9503121}].

\bibitem{Frolov:2005ty} S.~A.~Frolov, R.~Roiban and A.~A.~Tseytlin,
``Gauge-string duality for superconformal deformations of $\mathcal{N}=4$ super Yang-Mills theory,''
JHEP {\bf 0507}, 045 (2005)
[\arxivlink{hep-th/0503192}].

\bibitem{Frolov:2005dj} S.~Frolov,
``Lax pair for strings in Lunin-Maldacena background,''
JHEP {\bf 0505}, 069 (2005)
[\arxivlink{hep-th/0503201}].

\bibitem{Alday:2005ww} L.~F.~Alday, G.~Arutyunov and S.~Frolov,
``Green-Schwarz strings in TsT-transformed backgrounds,''
JHEP {\bf 0606}, 018 (2006)
[\arxivlink{hep-th/0512253}].

\bibitem{Roiban:2003dw} R.~Roiban,
``On spin chains and field theories,''
JHEP {\bf 0409} (2004) 023
[\arxivlink{hep-th/0312218}].

\bibitem{Berenstein:2004ys} D.~Berenstein and S.~A.~Cherkis,
``Deformations of $\mathcal{N} = 4$ SYM and integrable spin chain models,''
Nucl.\ Phys.\ B {\bf 702} (2004) 49
[\arxivlink{hep-th/0405215}].

\bibitem{Beisert:2005if} N.~Beisert and R.~Roiban,
``Beauty and the twist: The Bethe ansatz for twisted N=4 SYM,''
JHEP {\bf 0508} (2005) 039
[\arxivlink{hep-th/0505187}].

\bibitem{Fokken:2013aea} J.~Fokken, C.~Sieg and M.~Wilhelm,
``Non-conformality of $\gamma_i$-deformed $\mathcal{N} = 4$ SYM theory,''
J.\ Phys.\ A {\bf 47} (2014) 455401
[\arxivlink{1308.4420}].

\bibitem{Spradlin:2005sv} M.~Spradlin, T.~Takayanagi and A.~Volovich,
``String theory in beta deformed spacetimes,''
JHEP {\bf 0511} (2005) 039
[\arxivlink{hep-th/0509036}].

\bibitem{Frolov:2005iq} S.~A.~Frolov, R.~Roiban and A.~A.~Tseytlin,
`Gauge-string duality for (non)supersymmetric deformations of $\mathcal{N} = 4$ super Yang-Mills theory,''
Nucl.\ Phys.\ B {\bf 731} (2005) 1
[\arxivlink{hep-th/0507021}].

\bibitem{Bergman:2000cw} A.~Bergman and O.~J.~Ganor,
``Dipoles, twists and noncommutative gauge theory,''
JHEP {\bf 0010} (2000) 018
[\arxivlink{hep-th/0008030}].

\bibitem{Bergman:2001rw} A.~Bergman, K.~Dasgupta, O.~J.~Ganor, J.~L.~Karczmarek and G.~Rajesh,
``Nonlocal field theories and their gravity duals,''
Phys.\ Rev.\ D {\bf 65} (2002) 066005
[\arxivlink{hep-th/0103090}].

\bibitem{Borsato:2016ose} R.~Borsato and L.~Wulff,
``Target space supergeometry of $\eta$ and $\lambda$-deformed strings,''
JHEP {\bf 1610} (2016) 045
[\arxivlink{1608.03570}].

\bibitem{Arutyunov:2009ga} G.~Arutyunov and S.~Frolov,
``Foundations of the $AdS_5 \times S^5$ Superstring. Part I,''
J.\ Phys.\ A {\bf 42} (2009) 254003
[\arxivlink{0901.4937}].

\bibitem{Eguchi:1980jx} T.~Eguchi, P.~B.~Gilkey and A.~J.~Hanson,
``Gravitation, Gauge Theories and Differential Geometry,''
Phys.\ Rept.\ {\bf 66} (1980) 213.

\bibitem{Hoare:2016hwh} B.~Hoare and S.~J.~van Tongeren,
``On jordanian deformations of $AdS_5$ and supergravity,''
J.\ Phys.\ A {\bf 49} (2016) no.43, 434006
[\arxivlink{1605.03554}].

\bibitem{Kawaguchi:2014fca} I.~Kawaguchi, T.~Matsumoto and K.~Yoshida,
``A Jordanian deformation of AdS space in type IIB supergravity,''
JHEP {\bf 1406} (2014) 146
[\arxivlink{1402.6147}].

\bibitem{Orlando:2016qqu} D.~Orlando, S.~Reffert, J.~i.~Sakamoto and K.~Yoshida,
``Generalized type IIB supergravity equations and non-abelian classical $r$-matrices,''
J.\ Phys.\ A {\bf 49} (2016) no.44, 445403
[\arxivlink{1607.00795}].


\end{thebibliography}
\end{document}